\documentclass[]{aastex63}

\usepackage{xcolor}
\usepackage{soul}
\usepackage{graphicx}

\received{December 7, 2020}
\revised{January 27, 2021}
\accepted{February 12, 2021}

\submitjournal{ApJ}

\shorttitle{Energy Fraction in Supra-thermal and Energetic Particles at Interplanetary Shocks}
\shortauthors{David et. al.}
\graphicspath{{./}{figures/}}

\begin{document}

\title{In-situ Measurement of the Energy Fraction in Supra-thermal and Energetic Particles at ACE, Wind, and PSP Interplanetary Shocks}

\author[0000-0003-1713-9466]{Liam David}
\affiliation{Lunar \& Planetary Laboratory, University of Arizona, Tucson, AZ 85721, USA}

\author[0000-0002-5456-4771]{Federico Fraschetti}
\affiliation{Center for Astrophysics $|$ Harvard \& Smithsonian, Cambridge, MA, 02138, USA}
\affiliation{Lunar \& Planetary Laboratory, University of Arizona, Tucson, AZ 85721, USA}

\author[0000-0002-0850-4233]{Joe Giacalone}
\affiliation{Lunar \& Planetary Laboratory, University of Arizona, Tucson, AZ 85721, USA}

\author[0000-0002-7388-173X]{Robert F.\,Wimmer-Schweingruber}
\affiliation{Institute of Experimental and Applied Physics, Kiel University, Kiel, Germany}
\affiliation{National Space Science Center, Chinese Academy of Sciences, Beijing, China}

\author{Lars Berger}
\affiliation{Institute of Experimental and Applied Physics, Kiel University, Kiel, Germany}

\author[0000-0002-3176-8704]{David Lario}
\affiliation{Heliophysics Science Division, NASA Goddard Space Flight Center, Greenbelt, MD, USA}
\begin{abstract}
The acceleration of charged particles by interplanetary shocks (IPs) can drain a non-negligible fraction of the plasma pressure. In this study, we have selected 17 IPs observed {\it in-situ} at $1\,\text{au}$ by the Advanced Composition Explorer (ACE) and the Wind spacecraft, and 1 shock at $0.8\,\text{au}$ observed by Parker Solar Probe (PSP). We have calculated the time-dependent partial pressure of supra-thermal and energetic particles (smaller and greater than $50\,\text{keV}$ for protons and $30\,\text{keV}$ for electrons, respectively) in both the upstream and downstream regions. The particle fluxes were averaged for 1 hour before and 1 hour after the shock time to remove short time scale effects. Using the MHD Rankine-Hugoniot jump conditions, we find that the fraction of the total upstream energy flux transferred to supra-thermal and energetic downstream particles is typically $\lesssim\!16\%$, in agreement with previous observations and simulations. Notably, by accounting for errors on all measured shock parameters, we have found that for any given fast magnetosonic Mach number, $M_{f}\!<7$, the angle between the shock normal and average upstream magnetic field, $\theta_{Bn}$, is not correlated with the energetic particle pressure; in particular, the partial pressure of energized particles does not decrease for $\theta_{Bn} \gtrsim 45^\circ$. The downstream electron-to-proton energy ratio in the range $\gtrsim\!140\,\text{eV}$ for electrons and $\gtrsim\!70\,\text{keV}$ for protons exceeds the expected $\sim\!1\%$ and nears equipartition ($>\!0.1$) for the Wind events.
\end{abstract}


\section{Introduction}\label{sec:intro}
Shock waves are ubiquitous in space and astrophysical plasmas \citep{Treumann:09}. A shock converts incoming kinetic energy into downstream thermal and magnetic energy with a certain, usually small, fraction in energized charged particles \citep[e.g.,][]{Drury:83}. In the late 1970's \cite[e.g.][]{Gosling_etal_1981} evidence of direct extraction of accelerated particles from the thermal populations was provided by combining International Sun-Earth Explorers-2 (ISEE-2) and ISEE-3 observations. However, only a handful of studies have investigated in-situ measurements of the energy fraction in the energized particles. The contribution of supra-thermal and energetic ions ($\gtrsim 10\,\text{keV}$) and electrons ($\gtrsim 100\,\text{eV}$) to the energy balance at interplanetary shocks (IPs) has been quantified only indirectly and in only a few instances. \cite{Mewaldt.etal:08} estimated that in large coronal mass ejections (CMEs, inferred kinetic energy $> 10^{31}$ erg), the fraction of the bulk kinetic energy transferred into particles presumably accelerated by the shocks driving the CMEs can attain $10-20\%$. Likewise, from a comparative estimate of the energy fractions in the magnetic field, the flare-associated energetic particles, and the CME bulk motion of $38$ large solar events, \cite{Emslie.etal:12} concluded that between $0.4$\% and $20\%$ of the CME kinetic energy in the solar wind frame is stored in solar energetic particles accelerated by the associated IPs. From a multi-spacecraft study of 10 solar energetic particle events, a comparable fraction was obtained \citep{Aschwanden.etal:17}, under the assumption that particles are accelerated at CME-driven shocks. In some cases, the transfer of a significant fraction of the upstream ram pressure into energized charged particles causes their partial pressure (i.e. the kinetic pressure of the supra-thermal and energetic particles combined) to equal or exceed the magnetic field pressure, an effect observed by \cite{Lario.Decker:2002,Marhavilas.Sarris:11,Russell.etal:2013,Lario.etal:15}. However, to our knowledge, {\it in-situ} measurements of the energy content in energized particles at shocks are lacking.

The observations described above suggest that the effect of energized particles on the energy balance at shocks cannot be ignored and that the modeling of shock secular evolution must account for it. In addition, observational studies of the energized electron-to-ion energy ratio have not been extensively pursued. IPs are unique among shocks since the energy fraction in energized particles can be measured accurately and directly; for extra-solar shocks, the energy partition can only be inferred indirectly and relies on models of the radiation emission mechanism.

An analysis of 258 Wind IPs \citep{Reames:12} found that a high $\theta_{Bn}$, along with a high shock speed and compression ratio, is correlated with events showing large particle acceleration signatures in $1-10\,\text{MeV/nucl}$ He ions. This result is consistent with the long-known result that shocks with $\theta_{Bn}>45^\circ$ are rapid accelerators \citep{Jokipii:82,Jokipii:87}. The acceleration efficiency at quasi-perpendicular shocks can exceed that at quasi-parallel shocks, provided there is sufficient pre-existing large-scale magnetic turbulence present in the upstream region \citep{Giacalone_2005a,Giacalone_2005b,Fraschetti.Giacalone:15}. However, the error estimate on the shock parameters plays a pivotal role in any robust conclusion on the efficiency of particle acceleration at the shock.

Combining observations from the Advanced Composition Explorer (ACE) and the Wind spacecraft for a number of IPs, \cite{Lario.etal:19} showed that the upstream energy spectrum starts deviating from thermal and exhibits a significant supra-thermal energized ion component within only a few minutes of the shock. This result suggests that the particle partial pressure can rapidly vary as the shock approaches the spacecraft. Thus, the time evolution of the particle pressure needs to be monitored beginning several hours before the shock to identify its role in the acceleration process.

Early-on hybrid simulations (kinetic ions/fluid electrons) also found that the downstream energetic ion partial pressure flux is at most $10-25\%$ of the upstream ram energy flux \citep{Giacalone.etal:97}. Moreover, such a fraction was found to be roughly independent of the magnetic obliquity for the quasi-parallel shocks analyzed ($\theta_{Bn}\!\leq\! 30^{\circ}$, as larger values of $\theta_{Bn}$ were not considered therein) and also independent of Alfv\'en Mach numbers $\geq\!6$ for a fixed sonic Mach number \citep{Giacalone.etal:97}. Monte Carlo simulations modeling particle acceleration at the terrestrial bow shock \citep{Ellison.Moebius:87} measured by the Active Magnetospheric Particle Tracer Explorers/Ion Release Module (AMPTE/IRM) found that particles with energy per charge $>\!10\,\text{keV/Q}$ drain on the order of a few percent of the upstream ram energy flux. A comparison of Monte Carlo and 1D hybrid simulations \citep{Ellison.etal:1993} yielded close agreement for proton and heavy ion distribution functions between both approaches, concluding that the minimum energy fraction in $>10\,\text{keV}$ protons is $\sim\!10\%$.

Comparable energy fractions seem to be drained into energized particles at other types of shocks. Upon crossing the solar wind termination shock, Voyager 2 measured the temperature of the shock-heated plasma to be roughly $10$ times smaller than expected from the MHD jump conditions since a large fraction of the solar wind energy is likely transferred to the pick-up ions \citep{Richardson.etal08}. The partial pressure of ions $>28\,\text{keV}$ was measured to be comparable to, or exceed, that of the thermal ion population, and was about twice as large as the magnetic field pressure \citep{Decker.etal:2008}. At Earth's bow shock, at least 15\% and as much as 20\% of the incident solar wind energy flux is converted into energetic particles above $\sim\!10\,\text{keV/Q}$ \citep{Ellison.etal:1990}. \cite{Slane.etal:2014} inferred that a non-negligible fraction of the bulk kinetic energy ($16\%$) is converted into accelerated particles in the {\it Tycho}'s supernova remnant (SNR) shock by using a hydrodynamic model of the broadband spectrum (from radio to multi-TeV radiation). A comparable fraction of accelerated protons to the total initial kinetic energy was found for the two extragalactic supernova remnants N132D \citep{Bamba.etal:18} and 0509-67.5 \citep{Hovey.etal:15}.

In this paper, we consider a number of IPs at $1\,\text{au}$ with accurately determined shock parameters; such a selection allows unambiguous conclusions to be drawn for the first time on the efficiency on the particle energization (from supra-thermal up to ~100s MeV) over a broad range of magnetic obliquity and Mach number. In particular, we combine, for a sample of $17$ IPs, data sets for the thermal plasma momenta and for the supra-thermal/high-energy tail of the proton and electron energy spectra to determine the fraction of the shock energy density that is transferred into the non-Maxwellian particle populations for a broad range of $\theta_{Bn}$ and the fast magnetosonic Mach number $M_{f}$. The distinction between supra-thermal ($\lesssim\!50\,\text{keV}$ for protons, and $\lesssim\!30\,\text{keV}$ for electrons) and energetic ($\gtrsim\!50\,\text{keV}$ for protons, and $\gtrsim\!30\,\text{keV}$ for electrons) particles is herein made according to the nominal energy ranges of the instruments used in this analysis. Since a single integral of energy spectrum is calculated for each particle population over the entire available energy range, the distinction between supra-thermal and energetic particles has no physical effect. We find that, for any given interval of $M_{f}$, the relative energy spent in energizing particles does not decrease with $\theta_{Bn}$ and increases with $M_f$. We also include one of the first IPs observed {\it in-situ} by Parker Solar Probe (PSP) at $0.8\,\text{au}$.
 
Section \ref{sect:data} details our data sources, gives an overview of the detectors on ACE and Wind used in this work, outlines our shock selection criteria, and describes all data pre-processing steps. In Section \ref{sect:analysis} we describe our analysis methods, including coordinate transformations, determination of energized particle properties, and error propagation. Section \ref{sect:results} outlines our results. We characterize the overall energy conservation, the fluxes of energized protons and electrons, and the trends within a six-hour time window of each shock. Section \ref{sect:discussion} discusses known sources of error and broader implications of this study. Section \ref{sect:conclusions} summarizes the main conclusions of this work and discusses avenues for future research.

\section{Data sampling \label{sect:data}}
\subsection{Data Sources}
The Center for Astrophysics $|$ Harvard \& Smithsonian (CfA) shocks database, used in the present study, catalogues IPs observed by the ACE and Wind spacecraft, providing shock parameters including the time, speed, and shock normal direction, as well as plasma parameters such as proton temperatures, proton densities, and magnetic fields both upstream and downstream of the shock using the RH08 method \citep{Szabo:94,Koval.Szabo:08}. The RH08 method determines the shock speed and the shock normal direction independently using a multi-parameter optimization approach, thereby demonstrating the uniqueness of the solution. For dynamically varying quantities such as the magnetic field, the database uses adaptive averaging within roughly $\pm20$ minutes of the shock time both upstream and downstream.

\subsection{Shock Selection}
We selected 8 fast-forward shocks observed {\it in-situ} by ACE and 9 by Wind between 1997 and 2013 which are listed in Tables \ref{shocktable:large} and \ref{shocktable:small}. These tables include all asymptotic ($\sim\!20$ minutes before and after the shock passage) plasma parameters relevant to our analysis. In particular, we provide $M_f$, the shock speed in the spacecraft frame $V_{sh}$, $\textbf{V}_{up}\cdot\hat{\textbf{n}}$, and $\textbf{V}_{down}\cdot\hat{\textbf{n}}$ in the spacecraft reference frame (where $\textbf{n}$ is the unit vector normal to the shock surface and $\textbf{V}_{up}$ and $\textbf{V}_{down}$ are the upstream and downstream flow velocities, respectively), the thermal proton densities $n_{up}$ and $n_{down}$, and temperatures $T_{up}$ and $T_{down}$, $\theta_{Bn}$, and the magnetic energy densities $\textbf{B}_{up}^2/8\pi$ and $\textbf{B}_{down}^2/8\pi$. Since errors on the shock speed and normal direction have the greatest effect on our final uncertainties, these were our most limiting selection criteria. A compression ratio greater than two was required, as was a fast magnetosonic Mach number $M_{f}\gtrsim1.5$. We chose $M_{f}$ rather than the Alfv\'en Mach number $M_A$ since the former depends on both the sound speed and Alfv\'en speed, thus better capturing the characteristics of the background plasma environment. In order to examine the relation between $M_{f}$ and $\theta_{Bn}$, we selected shocks with a range of values: the Mach numbers ranged from $1.49$ to $6.21$, characteristic of IPs, and $\theta_{Bn}$ spanned from $19.2^\circ$ to $89.8^\circ$ (see Table \ref{table:1}). Only a small fraction of the intensity profiles of energized protons show a discernible quasi-exponential pre-shock rise, with small-amplitude fluctuations, followed by a flat top in the downstream plasma region, as predicted in the linear Diffusive Shock Acceleration (DSA) scenario \citep{Parker:65}. In order to minimize the error estimates in the energy fluxes, we thoroughly scanned the spacecraft archives; several shocks were included with particle intensity profiles deviating from the DSA prediction (see \cite{Lario.etal:03} for a shock classification scheme of the ACE archive for years 1999-2003 based on $\sim\!47\,\text{keV}$ ion and $\sim\!38\,\text{keV}$ electron intensity-time profiles).

PSP was crossed by a CME-driven IPs on 2020 November 30, at a distance of $0.81\,\text{au}$ from the Sun, and recorded one of the highest energetic particle intensities observed thus far in this mission \citep{Mitchell.etal:2021,Giacalone.etal:21,Lario.etal:2021}. The shock occurred at a time when PSP was near its farthest point from the Sun in its seventh orbit and close to $1\,\text{au}$ like the other measurements presented in our study; thus, it has been included in this analysis. However, due to the poor quality of data available for this event, we excluded it from our final shock statistics. Special steps that differ from the ACE/Wind shocks are discussed below.

\newpage

\subsection{Energized Particle Fluxes}
\subsubsection{ACE Particle Data}\label{ACE Particle Data}
\begin{figure}
    \centering
    \plotone{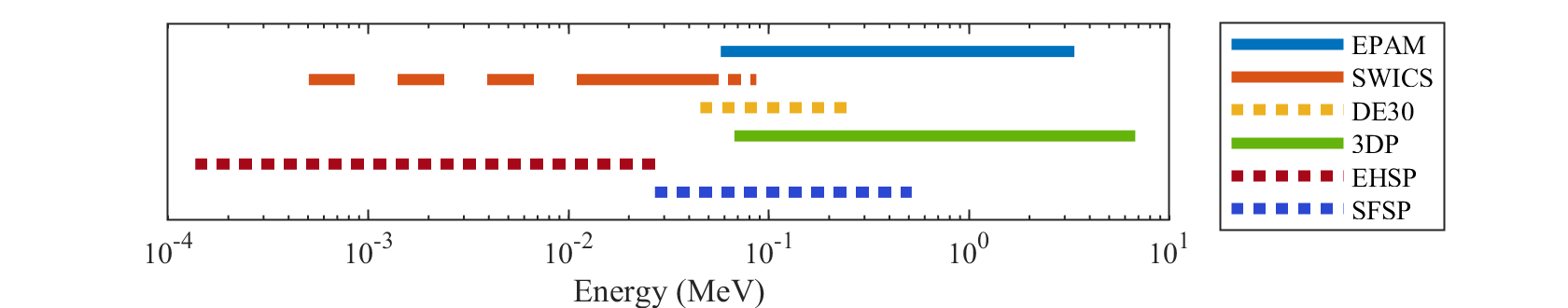}
    \caption{Instrument energy ranges used in this study for both protons (solid) and electrons (dotted). The lowest energies of SWICS are dashed, indicating the range below the DSA cutoff for all shocks. The highest four SWICS energy bins, which were discarded to eliminate overlap with EPAM, are dotted.}
    \label{fig:energy_ranges}
\end{figure}

Proton data for the ACE shocks were retrieved from the Electron, Proton, and Alpha Monitor (EPAM) through the CalTech Level 2 Database; \footnote{\url{http://www.srl.caltech.edu/ACE/ASC/level2/lvl2DATA_EPAM.html}} the Solar Wind Ion Composition Spectrometer (SWICS) instrument data were also gathered. EPAM contains a Low Energy Magnetic Spectrometer, designated LEMS120, that points $120^{\circ}$ off the spacecraft spin axis, where the ACE spin axis points mostly toward the Sun within $\pm20^\circ$. It records ion intensities between $47.0\,\text{keV}$ and $4.80\,\text{MeV}$ in 8 energy bins during the 12 seconds of the spacecraft spin period \citep{Gold.etal:98}. This energy range, and those of all instruments described below, are displayed in Figure \ref{fig:energy_ranges}, and fluxes were converted into units of $\text{cm}^{-2}\text{s}^{-1}\text{sr}^{-1}\text{eV}^{-1}$. SWICS is a linear time-of-flight (TOF) mass spectrometer with electrostatic deflection, allowing it to measure ion mass, charge, and energy. A full energy spectrum covering $657\,\text{eV}-86.6\,\text{keV}$ in $58$ bins is collected every 12 minutes; each bin is integrated serially for 12 seconds \citep{Gloeckler.etal:98}. SWICS can disentangle the background thermal solar wind, supra-thermal protons, and heavier ions, and here we consider only proton observations. Since EPAM extends down to $47.0\,\text{keV}$ and has both a higher time resolution and smaller errors than SWICS, we removed the highest four SWICS energy bins, consisting of protons above $48.8\,\text{keV}$, to eliminate overlap. In the overlapping energy region, SWICS intensities computed assuming isotropic fluxes may exhibit a lower value (by a factor of $\sim\!2$) than the EPAM intensities averaged over the same time interval, implying a possible discrepancy among instrument calibrations.

The formulation of DSA applies only to sufficiently isotropic (i.e. sufficiently fast) particles in the local plasma frame \cite{Parker:65} that can be efficiently accelerated by undergoing frequent pitch-angle scattering. Hybrid simulations \citep[e.g.,][]{Giacalone.etal:97} suggest a reasonable cut-off of 10 times the plasma ram energy, namely $v\gtrsim3V_{sh}$, where $v$ is particle speed in the spacecraft frame and $V_{sh}$ is also measured in the spacecraft frame. Thus, SWICS data below the DSA cutoff were not included (see Figure \ref{fig:energy_ranges}).

Spin-averaged electron data for the ACE shocks were obtained from the Deflected Electrons (DE30) instrument, which points $30^\circ$ off the spin axis, through the NASA CDAWeb database.\footnote{\url{https://cdaweb.gsfc.nasa.gov/index.html/}} Four energy bins span $38\,\text{keV} - 315\,\text{keV}$ and the time resolution is 12 seconds. A magnetic deflector separates these electrons from the parent LEMS30 instrument, which are measured, although with an elevated instrumental background, by a totally depleted surface barrier silicon detector \citep{Gold.etal:98}.

\subsubsection{Wind Particle Data}
Omnidirectional proton flux data for the Wind shocks were obtained from the Three-Dimensional Plasma and Energetic Particle Investigation (3DP) instrument through the NASA CDAWeb database. The pertinent detector is a solid-state telescope with a $36^\circ\times20^\circ$ field of view $126^\circ$ off the spin axis which points perpendicular to the ecliptic plane. Nine ion energy bins span $67.3\,\text{keV}-6.76\,\text{MeV}$ and are integrated in parallel for 12-second samples.

We also considered proton fluxes in the energy range $400\,\text{eV}-19.1\,\text{keV}$ from the PESA-Hi (PH) data product, however only one or two of the highest PH energy bins lie beyond the DSA cutoff. Moreover, variations in the definition of the energy bins, as well as uncertainties in estimating the energy flux, led us to discard the PH energy range. Likewise, we did not include here Wind/SMS/STICS (SupraThermal Ion Composition Spectrometer) data \citep{Gloeckler.etal:1995} that covers the ion energy range $6-230\,\text{keV/e}$, because the reported fluxes\footnote{\url{https://spdf.gsfc.nasa.gov/pub/data/wind/sms/l2/stics/vdf_ascii/solarwind/}} for the selected events are vanishing for most of the angular sectors within $\pm 3$ hours of each shock. 

Wind 3DP also contains two electron data products: EHSP spans $136\,\text{eV}-27.6\,\text{keV}$ in 15 energy bins and SFSP covers $27.0\,\text{keV}-517\,\text{keV}$ in $7$ bins. Omnidirectional data with a $\sim\!24\,\text{s}$ resolution were obtained from the NASA CDAWeb database. EHSP data are compiled from the high-range Electron Electrostatic Analyzer (EH), which has a $360^\circ\times90^\circ$ field of view and works by electrostatic deflection and micro-channel plate detection. SFSP data are from the Solid State Foil Telescope (SF) instrument having a $180^\circ\times20^\circ$ field of view across five apertures. See \cite{Lin.etal:95} for a full description of these instruments. We did not compensate for the small overlap ($0.6\,\text{keV}$) between SFSP and EHSP since removing either the highest EHSP or lowest SFSP energy bin would leave a large energy range ($18.9\,\text{keV}-27.0\,\text{keV}$ or $27.6\,\text{keV}-40.1\,\text{keV}$, respectively) unaccounted for.

\newpage

\subsubsection{PSP Particle Data}
We obtained energized proton fluxes from the Integrated Science Investigation of the Sun (IS$\odot$IS), Energetic Particle Instruments (EPI-Lo and EPI-Hi), which measure ion energies in the ranges $\sim\!20\,\text{keV/nuc}-15\,\text{MeV}$ and $\sim\!1-200\,\text{MeV/nuc}$, respectively. 
EPI-Lo is a TOF mass spectrometer capable of particle species differentiation using a solid-state detector (SSD), and EPI-Hi uses ion-implanted SSDs and the $\text{d}E/\text{d}x$ versus energy method for both energy and species determination \citep{McComas.etal:2016}. For the PSP event that occurred on 2020 November 30, EPI-Hi fluxes are unavailable above $20\,\text{MeV}$ and before 2020 December 1 at 4:29:57 UT, corresponding to the first 10 hours after the shock passage. This event is added herein only to compare the order of magnitude of the PSP {\it in-situ} measurements for a particularly strong event at $0.8\,\text{au}$ with those at $1\,\text{au}$. Electrons are not included due to the significant uncertainties in the measurements of this event.

\section{Analysis Method \label{sect:analysis}}
\subsection{ACE and Wind Shocks}

For the ACE and Wind shocks, the upstream and downstream particle fluxes were found by averaging each energy bin for 1 hour. The interval upstream ended one minute before the shock crossing, and the interval downstream began 1 minute after the shock time to best match the CfA database plasma parameters. Ideally the analysis window would have to be reduced to $\lesssim\!20$ minutes to match the CfA plasma quantities, which are typically averaged over $10-20$ minutes within 20 minutes of the shock time. However, this choice would be in conflict with the lower SWICS time resolution (12 minutes), and 5 samples were deemed necessary to reduce the effects of fluctuations in the data. Therefore, we used 300-sample averages for EPAM, DE30, and 3DP, and 150-sample averages for EHSP and SFSP, to maintain a consistent 1 hour duration across measurements from the different instruments.

We are interested in the contribution of energized particles with speeds $v\gtrsim3V_{sh}$ to the enthalpy flux (partial pressure times the bulk flow speed). The maximum shock speed $V_{sh}$ in our dataset is $1,105.7\pm50.5\,\text{km/s}$, so the minimum energy cutoffs across the entire sample in the downstream medium for protons and electrons are $57.4\,\text{keV}$ and $31.3\,\text{eV}$, respectively. For the slowest shock ($V_{sh}=477.7\pm8.7\,\text{km/s}$), these cutoffs are reduced to $10.7\,\text{keV}$ and $5.84\,\text{eV}$. SWICS energy bins (minimum energy $657\,\text{eV}$) are below these thresholds and extend into the thermal part of the proton spectrum. As mentioned in Section \ref{ACE Particle Data}, for the energized particle contribution, we use only those SWICS fluxes above the energy cutoff, which was calculated individually for each shock. Notably, the nominal lower energies for the remaining low-energy instruments, DE30, 3DP, and SFSP, lie above the DSA cutoff, hence excluding parts of the energized particle populations. The unaccounted energy ranges involve electrons $\sim\!6\,\text{eV}-38\,\text{keV}$ for ACE, and both protons $\sim\!10.7\,\text{keV}-67.3\,\text{keV}$ and electrons $\sim\!6\,\text{eV}-138\,\text{eV}$ for Wind. Since particle fluxes increase with decreasing energy, the calculated downstream supra-thermal Wind proton partial pressures lack a possibly non-negligible contribution in the range $\sim\!10.7\,\text{keV}-67.3\,\text{keV}$. See Section \ref{sect:missing_energy} for further discussion.

A fraction of the bulk incoming kinetic energy is channeled into producing low-energy supra-thermal ion populations reflected after single or multiple shock encounters, i.e., field-aligned beam and gyrophase bunched, respectively, whose admixture was studied as a function of distance from the shock in 2D simulations by \cite{Savoini.etal:13}. The typical energy of both populations, as measured at IPs, can be as high as  $\sim\!20\,\text{keV}$ \citep[e.g.,][]{Kajdic.etal:17}, higher than the $\sim\!10\,\text{keV}$ measured at the Earth bow shock \citep[e.g.,][]{Meziane.etal:05}. Thus, such ion populations overlap only marginally with the energy range here analysed (protons $>\!10.7\,\text{keV}-57.4\,\text{keV}$ and electrons $>\!5.84\,\text{eV}-31.3\,\text{eV}$). Therefore, restricting the partial pressure of energized particles to only the downstream medium does not exclude the upstream reflected ion components, relevant only at lower energies. Finally, the partial pressure in the upstream medium requires disentangling the pristine incoming energized particles from the shock-reflected ones. The 3D velocity distribution necessary for this task is available only for Wind data and was not considered in this work.

The CfA database parameters, as well as the particle fluxes, are provided in the spacecraft reference frame, whereas the jump conditions are usually, and conveniently, expressed in the shock reference frame. The observed upstream and downstream solar wind velocities were converted to the shock frame using
\begin{equation}
    \textbf{v}_{up/down}^{(sh)}=\textbf{V}_{up/down}^{(sc)}-V_{sh}^{(sc)}\hat{\textbf{n}} \,.
\end{equation}
Subsequent equations will use identical super-scripting where $(sh)$ denotes the shock frame and $(sc)$ the spacecraft frame. 

The particle energies for all instruments were relativistically transformed into the shock frame. We used the energies given by the geometric means of the energy/momentum bins $E_s^{(sc)}$. Subscript $s$ denotes a quantity corresponding to a particle species (protons and electrons). Velocities in the spacecraft frame are defined as
\begin{equation}
    v_s^{(sc)}=c\sqrt{1-\left(\frac{E_p^{(sc)}}{m_sc^2}+1\right)^{-2}}\textbf{\,,}
\end{equation}
where $m_s$ is the rest mass of species $s$, and then are boosted using
\begin{equation}
    v_s^{(sh)}=\frac{v_s^{(sc)}-V_{sh}^{(sc)}}{1-\frac{v_s^{(sc)}V_{sh}^{(sc)}}{c^2}}\,.
\end{equation}
The shock-frame momenta and energies are
\begin{equation}
    p_s^{(sh)}=\gamma m_sv_s^{(sh)}
\end{equation}
\begin{equation}
    E_s^{(sh)}=(\gamma-1)m_sc^2\,,
\end{equation}
where $\gamma$ is the relativistic gamma factor involving $v_s^{(sh)}$. The fluxes are assumed, both upstream and downstream, to be isotropic, meaning the particle mass density and pressure are
\begin{equation}
    \rho_s=4\pi m_s\int_{p_{min}} ^{p_{max}} Fdp_s^{(sh)}
\end{equation}
\begin{equation}
    P_s=\frac{4}{3}\pi m_s\int_{E_{min}} ^{E_{max}} \gamma v_s^{(sh)}F dE_s^{(sh)}\,.
\end{equation}
Here the flux $F$ is measured in $cm^{-2}s^{-1}sr^{-1}eV^{-1}$. The integrals were approximated using the trapezoidal method.

The gas and particle specific enthalpies were calculated using
\begin{equation}
h_s=\frac{\Gamma}{\Gamma-1}\frac{P_s}{\rho_s} \,.
\label{eq:enthalpy}
\end{equation}
Here $\Gamma=5/3$ is the adiabatic index, which is set by our assumption of a monatomic gas, and $P_s$ and $\rho_s$ are the pressures and mass densities. The magnetic field was not transformed into the shock frame since the shocks are non-relativistic.

The Rankine-Hugoniot (RH) jump conditions express the conservation of mass, momentum, energy, and magnetic flux of a magnetized fluid heated by a shock. A coordinate system was constructed in the shock frame such that $\hat{\textbf{z}}=\hat{\textbf{n}}$. The $\hat{\textbf{x}}$ component was set parallel to the component of $\textbf{B}_{up}$ tangent to the shock surface and $\hat{\textbf{y}}$ completes the right-handed set. All RH jump conditions are referenced to this system, which is different for each shock. We decompose the magnetic fields and velocities as
\begin{equation}
    \textbf{B}_{up/down}=B_k\hat{\textbf{z}}+\textbf{B}_l
\end{equation}
\begin{equation}
    \textbf{v}_{up/down}^{(sh)}=v_k\hat{\textbf{z}}+\textbf{v}_l
\end{equation}
with the magnitudes $|\textbf{B}_{up/down}|=B$ and $|\textbf{v}_{up/down}^{(sh)}|=v$. The subscripts $l$ and $k$ indicate quantities tangent and normal to the shock surface, respectively.
The energy flux jump condition is
\begin{equation}
    F^{up}_{TOT}|_{h_s =0} - F^{down}_{TOT} = \left[v_k\left(\rho_gh_g+\frac{1}{2}\rho_gv^2+\sum_s \rho_sh_s\right)+\frac{1}{4\pi}\left(v_kB^2-B_k(\textbf{v}\cdot\textbf{B})\right)\right]|_{h_s^{up} =0}=0 \, ,
    \label{eq:RH_energy}
\end{equation}
where $F^{up/down}_{TOT}$ are the total fluxes of energy density upstream and downstream, respectively, $[\,]$ indicates the difference between upstream and downstream quantities, and the subscript $g$ denotes thermal gas quantities.\footnote{The downstream enthalpy flux of the energized particle species $s$ in Equation \ref{eq:enthalpy}, $h_s \rho_s v_k$, corresponds to the term $(5/2)P U$ in \cite{Giacalone.etal:97}} For later reference we define the kinetic energy flux $F_{KIN} = v_k\rho_gv^2$, the enthalpy flux $F_{THERM} = v_k\rho_gh_g$, and the magnetic energy fluxes $F_{MAG,1} = v_kB^2/(4\pi)$ and $F_{MAG,2} = -B_k(\mathbf{v}\cdot\mathbf{B})/(4\pi)$. The normal momentum jump condition is

\begin{equation}
    \left[\rho_gv_k^2+P_g+\sum_s P_s+\frac{B_l^2}{8\pi}\right]|_{h_s^{up} =0}=0 \, .
    \label{eq:RH_momentum}
\end{equation}
As mentioned above, the energized particle fluxes can rise steeply (quasi-exponentially) before the shock time. These particles originate downstream very close to the shock; some then diffuse upstream. Adding these particles to the upstream term of the jump conditions double-counts the energy/momentum of energetic particles, and misplaces into the upstream energy/momentum that likely originated downstream. Thus, only the downstream portion of the jump conditions includes $\sum_s\rho_sh_s$ and $\sum_sP_s$.

\subsection{PSP Shock}
The PSP shock on 2020 November 30 at 18:35 UT \citep{Mitchell.etal:2021} is not catalogued in the CfA database nor are pre-computed shock/background plasma parameters available. The $\theta_{Bn}$ is estimated to be $\sim 60^\circ$ \citep{Giacalone.etal:21}. Using polarization data from the Radio Frequency Spectrometer (RFS) of the FIELDS instrument on board PSP \citep{Bale.etal:2016}, upstream waves seem to be close to field-aligned, which would be consistent with electrostatic Langmuir waves \footnote{Dr. M. Pulupa, private communication}. The low frequency ($18.75\,\text{kHz}$) corresponds to a density of $\sim\!4.3\, \text{cm}^{-3}$. The spacecraft spatial orientation at the time of the shock passage made it impossible to accurately determine the shock speed and the downstream proton thermal speed using the Solar Wind Electrons Alphas \& Protons (SWEAP) instrument \citep{Kasper.etal:2016}; reasonable values to assume seem to be $\sim\!900\,\text{km/s}$ and $40\,\text{km/s}$, respectively. The PSP/FIELDS downstream magnetic field, which was obtained from the CDAWeb database, averaged over 10 minutes, is $\sim\!36.7\,\text{nT}$. We can infer the downstream number density by using the slopes of sections of the time-averaged particle energy spectra, assumed to be power-laws from EPI-Hi and EPI-Lo measurements. In particular, EPI-Lo fluxes were averaged for 1 hour after the shock passage, constituting 12 samples of 5-minute duration each, after the fluxes reached a clear plateau about 20 minutes after the shock time. The EPI-Hi data are unavailable for about 10 hours after the shock but still show enhancement over background. We averaged the first two available fluxes in each energy bin which corresponds to a period of two hours. In the assumption of linear DSA, the compression ratio is then
\begin{equation}
    \frac{n_{down}}{n_{up}}=\frac{q}{q-3} \,,
\end{equation}
where $q$ is the absolute value of the power-law index. Averaging the slopes of three regions of the spectra, we obtain a reasonable ratio of $3.1$. We compute a total downstream energy flux of $\sim\!9.2\,\text{erg/s/cm}^2$, roughly an order of magnitude larger than that for the ACE and Wind shocks (see results in Section \ref{sect:results}). However, with these data uncertainties we are unable to accurately verify the overall energy conservation.

\subsection{Error Propagation}
Particular attention has been paid to calculate the errors on each quantity appearing in the jump conditions (Equations \ref{eq:RH_energy} and \ref{eq:RH_momentum}), as well as on the fraction of the total energy in downstream supra-thermal and energetic particles. For each shock, we generated $10^5$ sample shocks with all parameters randomly perturbed according to a normal distribution having as mean and standard deviation the values provided in the CfA database. These shocks were individually analysed and the overall uncertainties on our results determined using the standard deviations. Therefore, the standard deviations calculated herein account for experimental sources of error.

\newpage

\section{Results} \label{sect:results}
\subsection{Energy Partition Among Plasma Components}
\begin{figure}
    \gridline{\fig{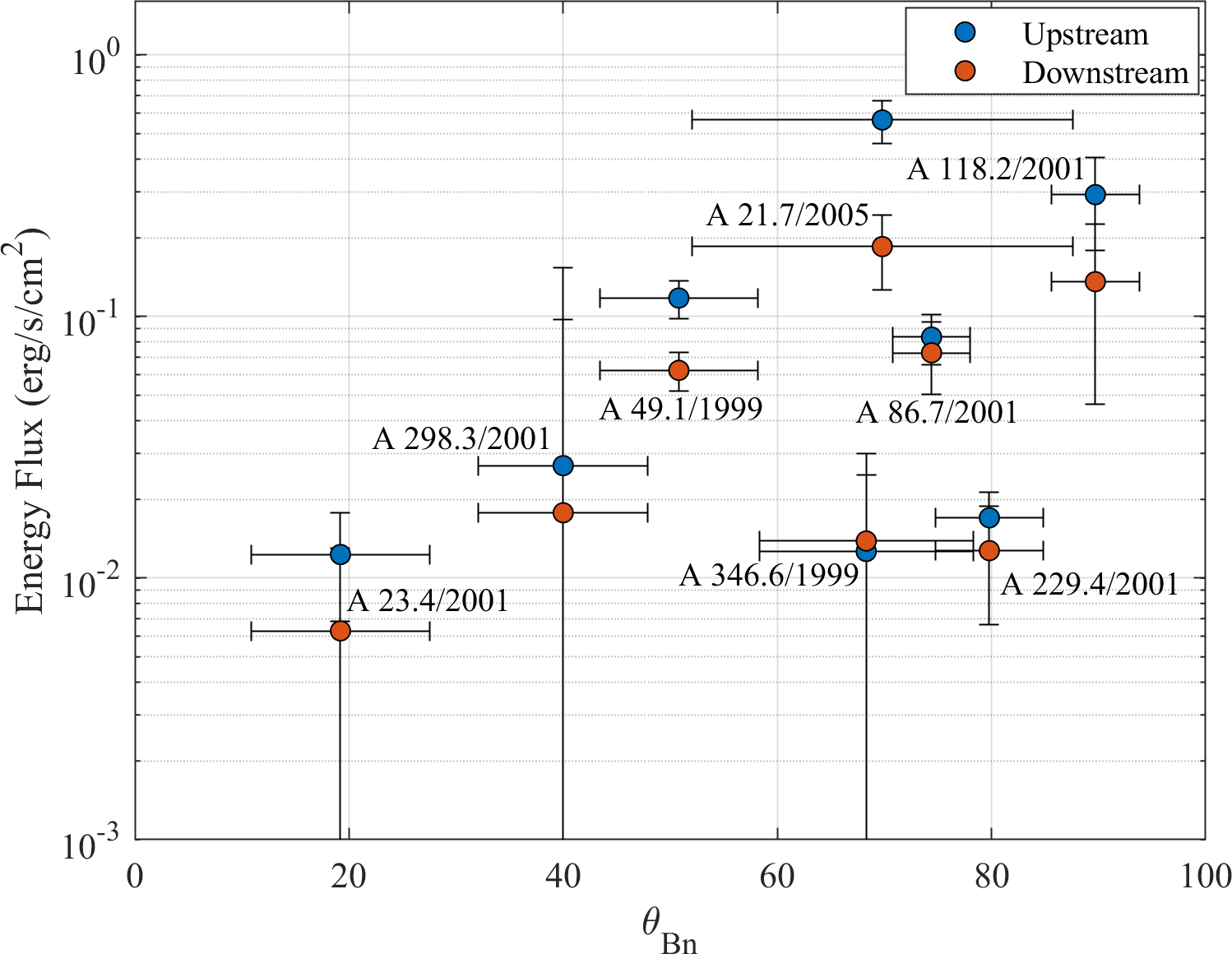}{0.48\textwidth}{}
              \fig{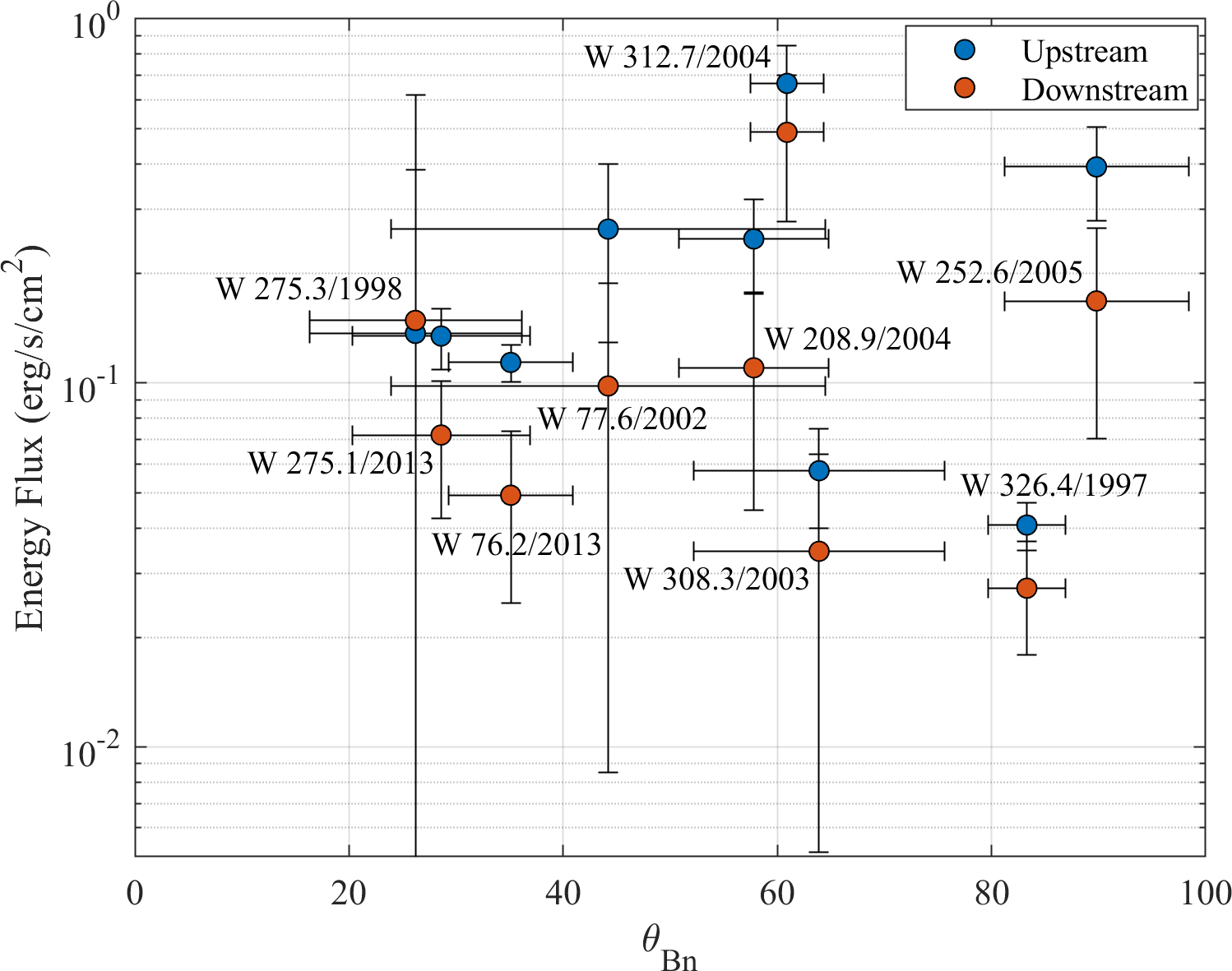}{0.48\textwidth}{}}
    \vspace{-0.5cm}
    \gridline{\fig{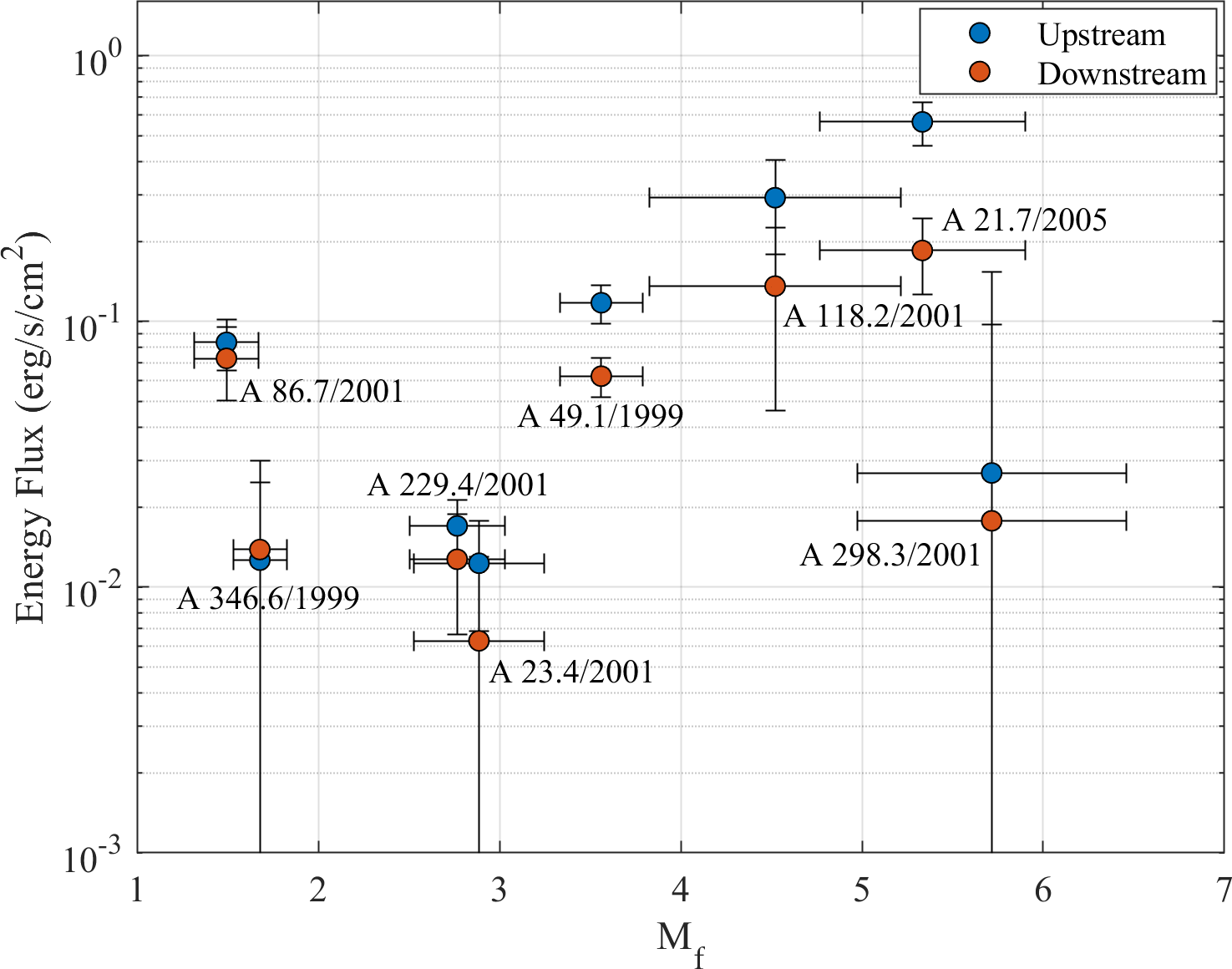}{0.48\textwidth}{}
              \fig{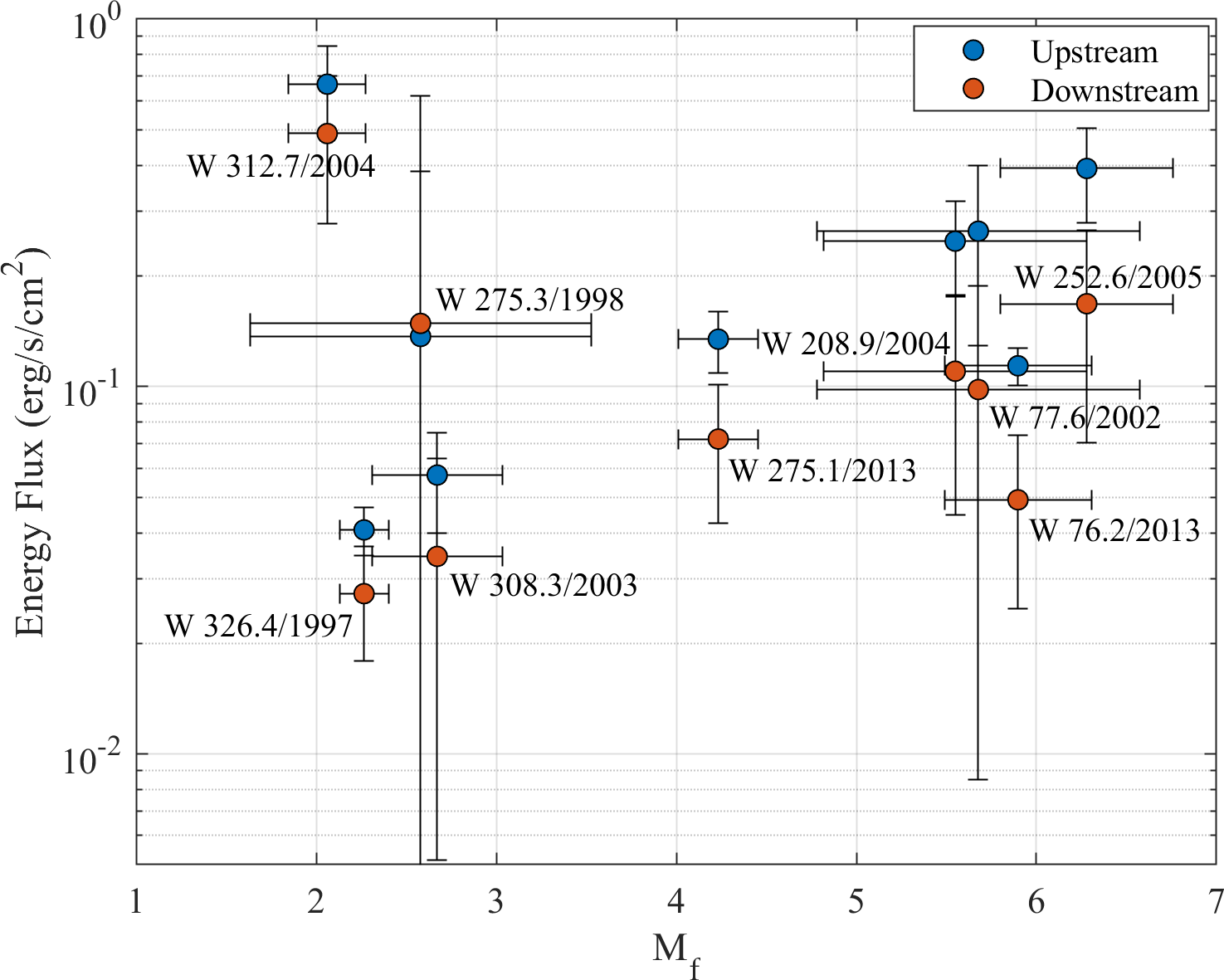}{0.48\textwidth}{}}
    \vspace{-0.5cm}
    \caption{Upstream (blue) and downstream (red) energy fluxes at each shock with $1\sigma$ error bars for a 1-hour pre- and post-shock interval. The left and right panels correspond to ACE and Wind, respectively. The top row is plotted as a function of $\theta_{Bn}$ and the bottom as a function of $M_{f}$. Energized particle fluxes are included only in the downstream values to avoid contamination from particles diffusing from downstream (see Section \ref{sect:results}). Particle fluxes are averaged for 1 hour beginning 1 minute after the shock time.}
    \label{fig:RH_energy}
\end{figure}

\begin{figure}
    \gridline{\fig{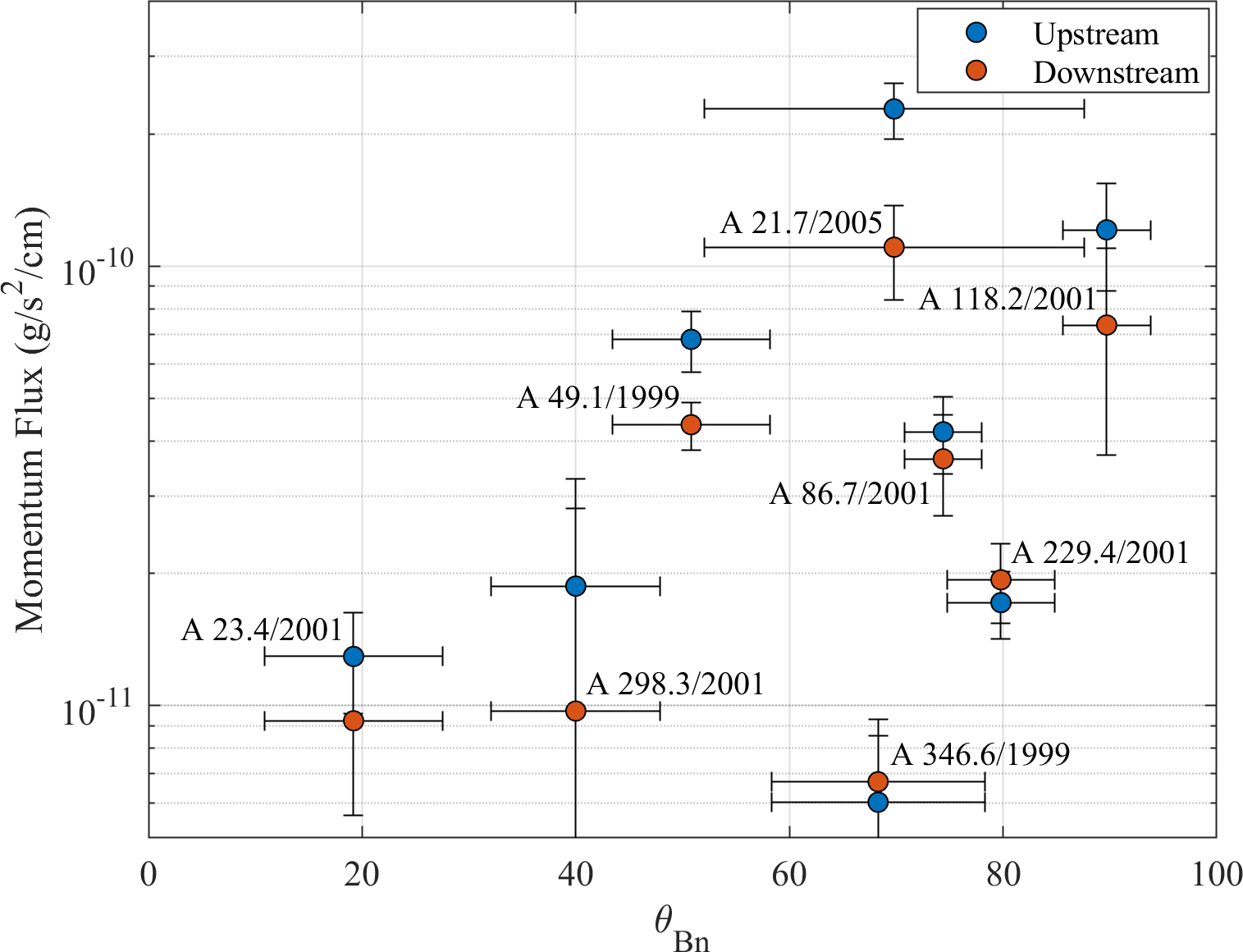}{0.48\textwidth}{}
              \fig{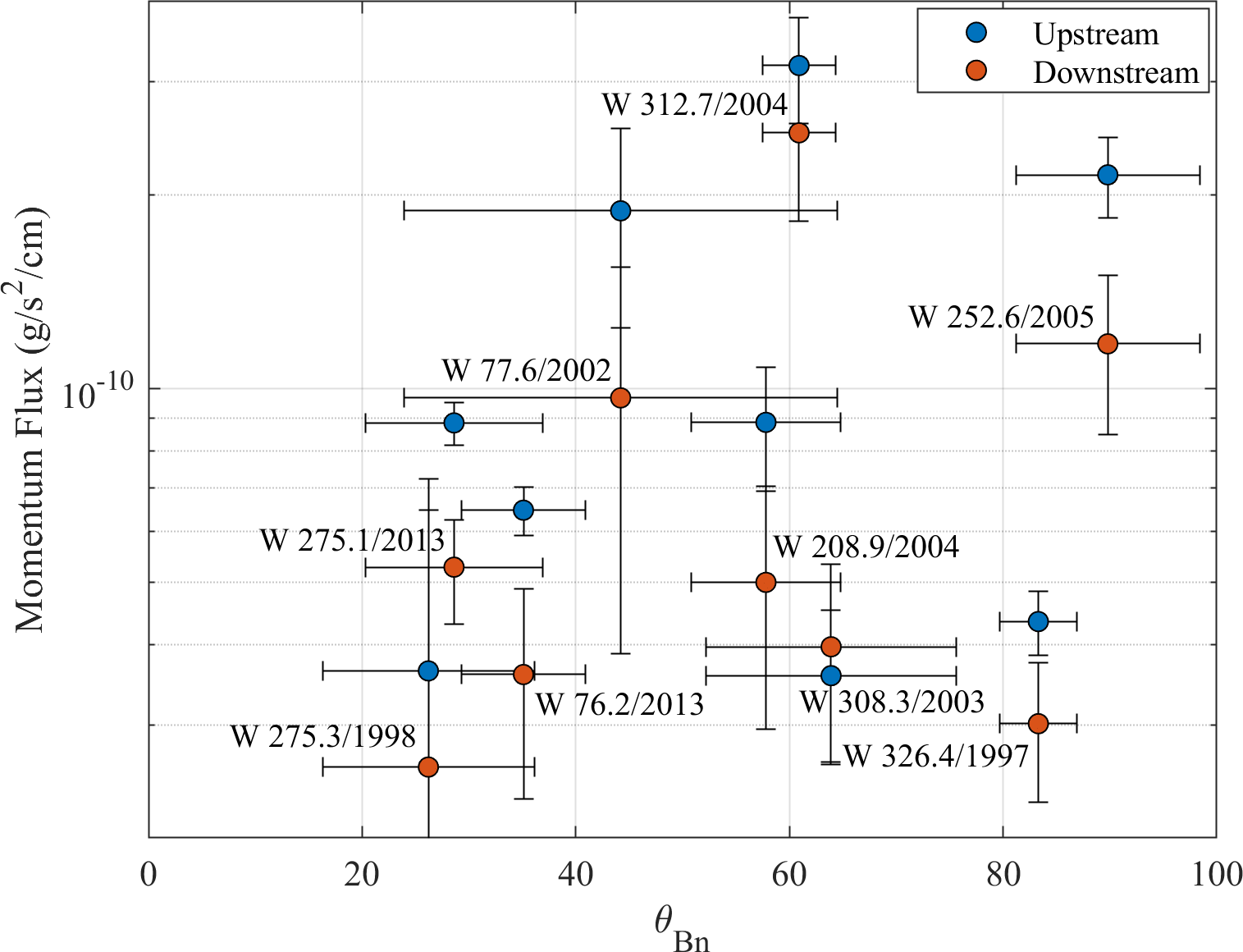}{0.48\textwidth}{}}
    \vspace{-0.5cm}
    \caption{Upstream (blue) and downstream (red) normal momentum fluxes at each shock as a function of $\theta_{Bn}$ with $1\sigma$ error bars. The left and right panels correspond to ACE and Wind, respectively. Energized particle energy fluxes are included only in the downstream region. One-hour averages offset by 1 minute from the shock times were used for particle flux determination.}
    \label{fig:RH_momentum}
\end{figure}

Figure \ref{fig:RH_energy} plots the energy flux upstream and downstream ($F^{up}_{TOT}|_{h_s=0}$ and $F^{down}_{TOT}$, Equation \ref{eq:RH_energy}) as a function of $\theta_{Bn}$ (top panels) and $M_f$ (bottom panels) for a 1-hour pre- and post-shock interval. Six of the eight ACE shocks (23/2001, 298/2001, 346/1999, 86/2001, 229/2001, and 118/2001) and five of the nine Wind shocks (326/1997, 275.3/1998, 77.6/2002, 312/2004, and 308/2003) show a matching of upstream and downstream energy fluxes within $1\sigma$ errors. Only ACE 21/2005 does not match within $2\sigma$. Figure \ref{fig:RH_momentum} shows the normal momentum RH jump conditions (Equation \ref{eq:RH_momentum}), exhibiting the same pattern: the shocks that match within $1\sigma$ in energy flux do likewise in momentum flux, except for Wind 326/1997.

Figure \ref{fig:energy_fractions} shows the values of the different energy flux components for each shock normalized to the total energy flux, in the respective region, upstream (left panel) and downstream (right panel). The upstream ram pressure (red), which dominates the energy flux, is efficiently converted downstream into other components by the shock, mostly magnetic (yellow) and thermal (blue). Wind events show an energized electron energy fraction (EHSP + SHSP, gray + lime) not negligible with respect to that of ions (dark red); the same cannot be concluded about the ACE events due to the large electron energy cutoff of DE30 data.

All ACE events (except 23.4/2001 and 298.3/2001, which have the smallest $\theta_{Bn}$) and all Wind events show a large magnetic energy fraction downstream, as well as a large fraction increase, suggesting that the field might undergo a downstream amplification beyond compression, possibly due to upstream density inhomogeneities \citep{Giacalone.Jokipii:07,Fraschetti:13}. Comparably large magnetic fractions are not found at the two quasi-parallel events. For the shocks ACE 298/2001, 346/1999, 86/2001, and the Wind shocks 252/2005, 326/1997 the term $F_{MAG,2}$ is negative, resulting in normalizations greater than one.

\subsection{Energetic Particle Fractions}
\begin{figure}
    \gridline{\fig{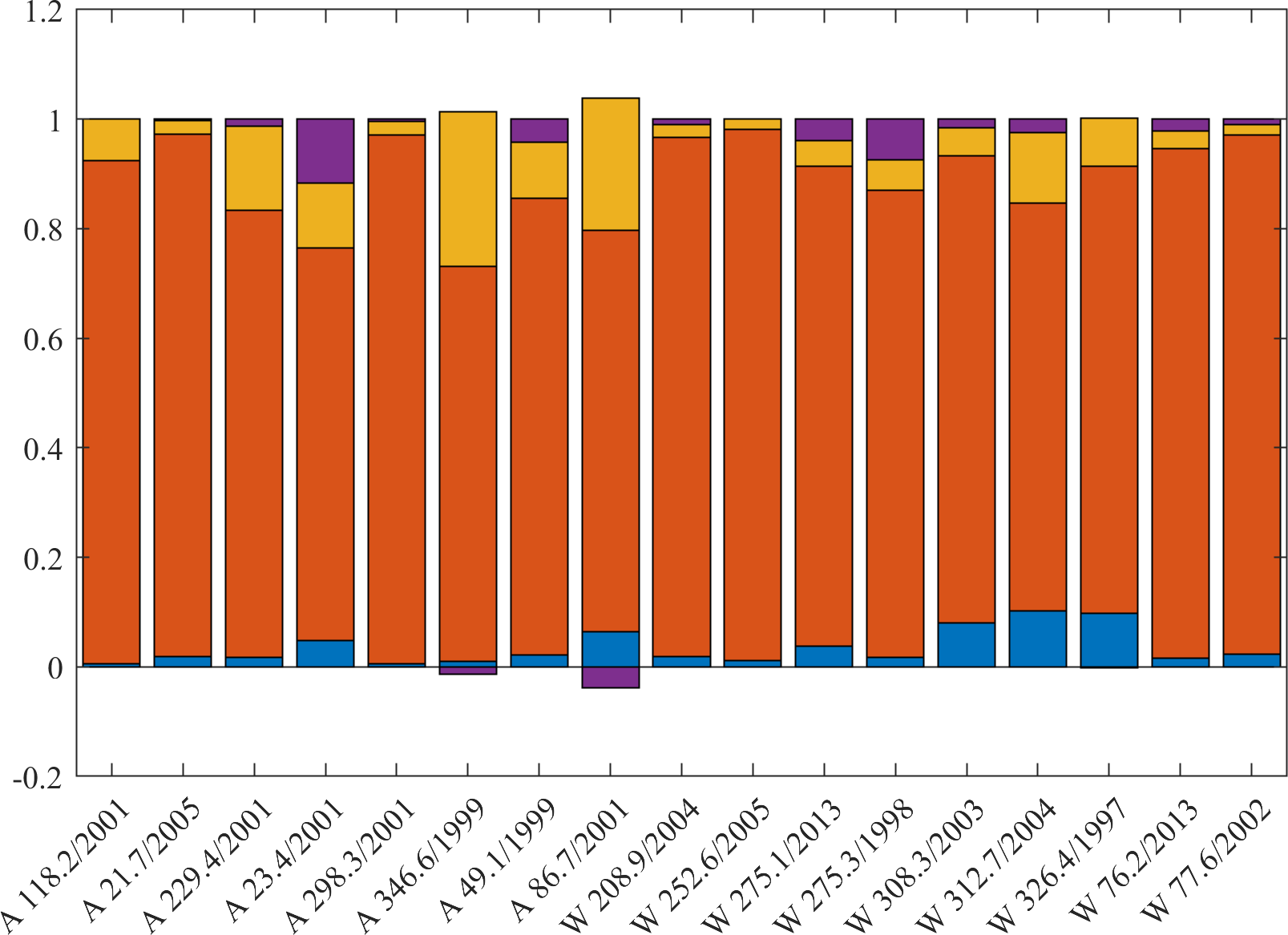}{0.45\textwidth}{}
              \fig{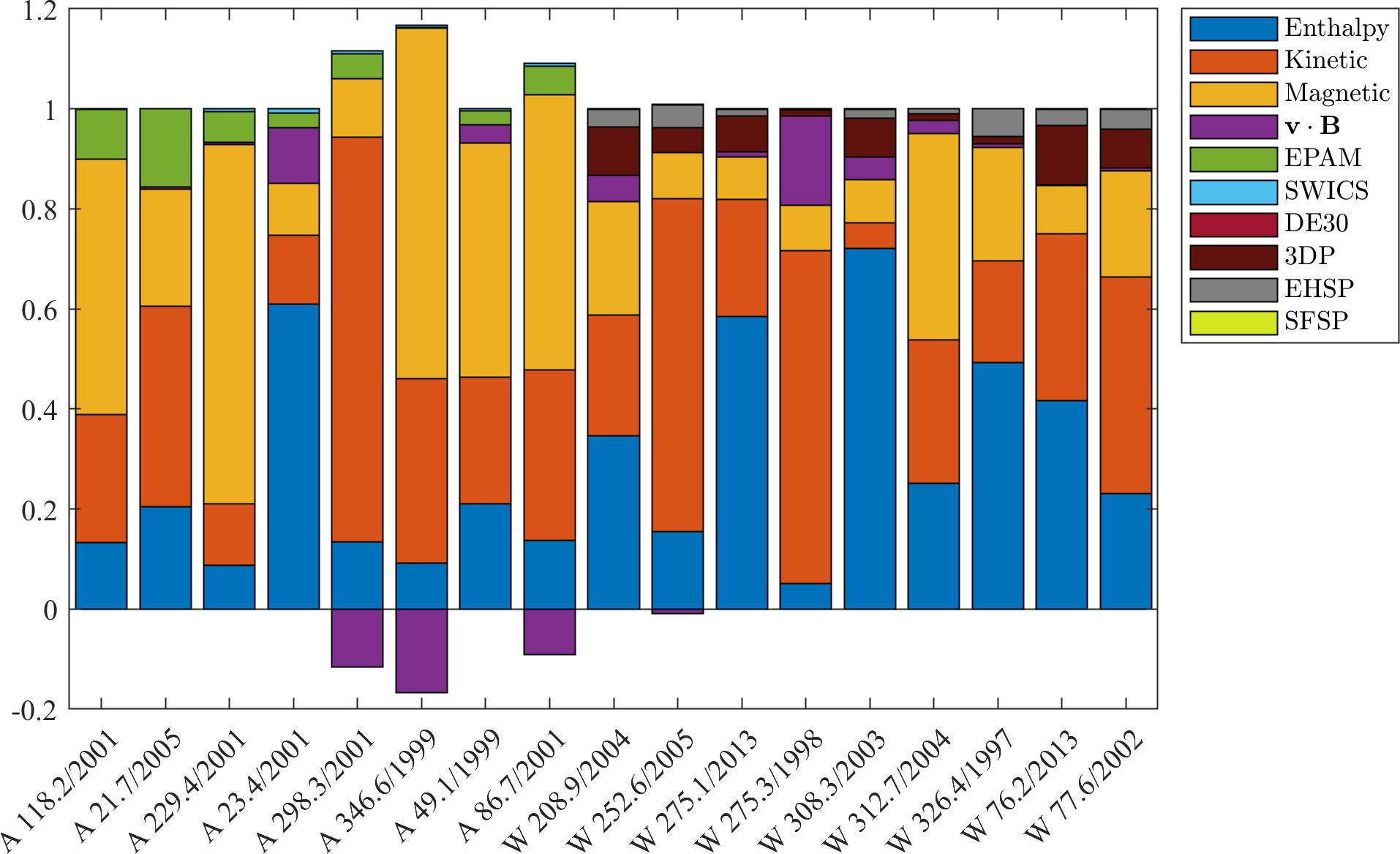}{0.535\textwidth}{}}
    \vspace{-0.5cm}
    \caption{By comparing, for the ACE and Wind shocks, the energy partition in the upstream (left panel) with the downstream (right panel), the efficient conversion of the shock kinetic energy ($F_{KIN}/F_{TOT}$, red) into the plasma enthalpy ($F_{THERM}/F_{TOT}$, blue), magnetic pressure ($F_{MAG,1}/F_{TOT}$ yellow), the $\mathbf{v}\cdot\mathbf{B}$ term ($F_{MAG,2}/F_{TOT}$, purple) and energized particles ($v_k\rho_sh_s/F_{TOT}$, all other colors) is apparent. The total energy flux has been normalized to 1 and the 1-minute offset bins for both upstream and downstream as described in Section \ref{sect:time_series} were used.}
    \label{fig:energy_fractions}
\end{figure}

\begin{figure}
    \gridline{\fig{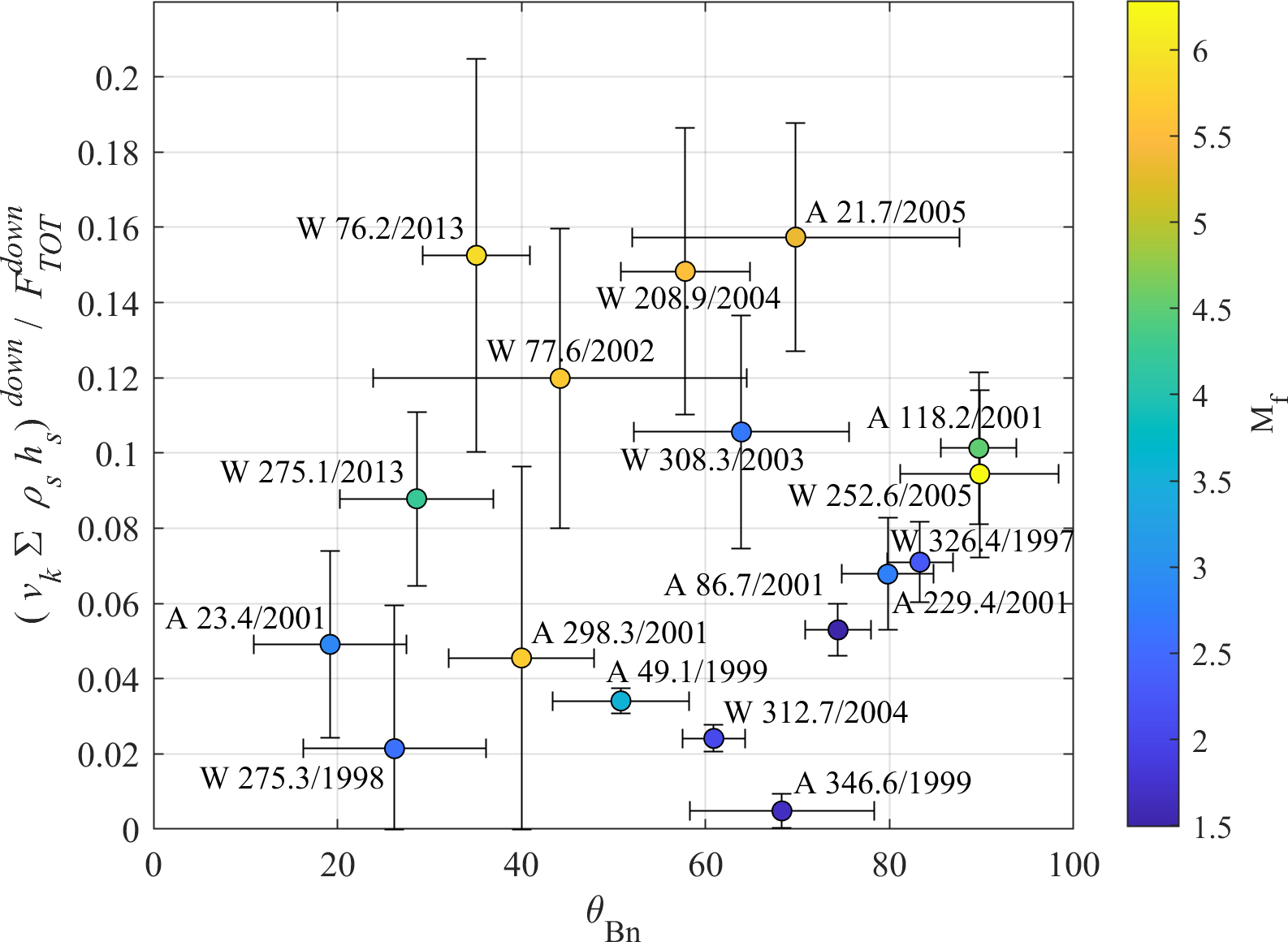}{0.49\textwidth}{}
              \fig{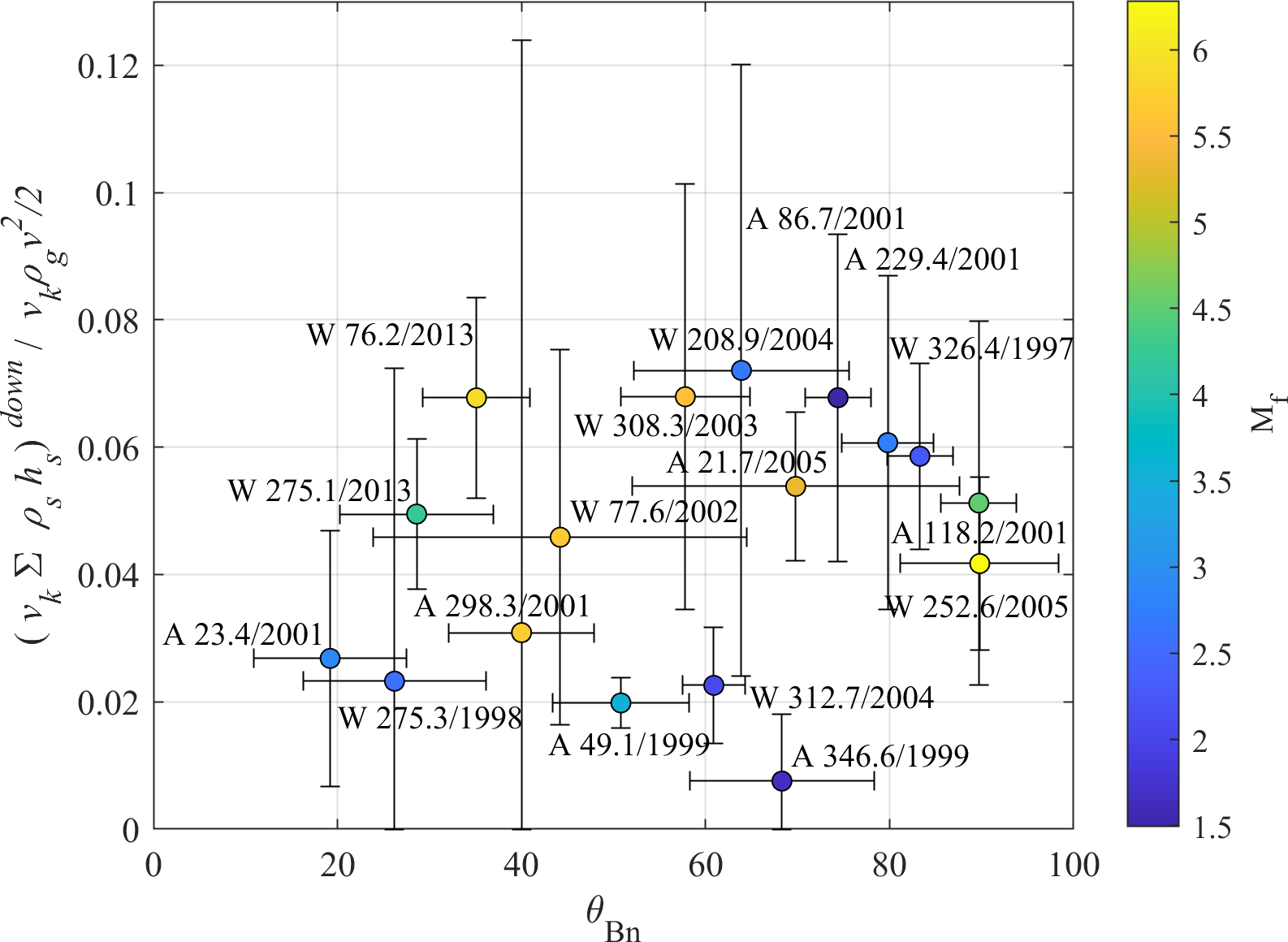}{0.49\textwidth}{}}
    \vspace{-0.5cm}
    \caption{\textbf{Left:} Ratio of downstream energy flux of energized particles (protons $+$ electrons, $(v_k\sum\rho_sh_s)^{down}$) to the total downstream energy flux ($F^{down}_{TOT}$) as a function of $\theta_{Bn}$ for both ACE and Wind. Spacecraft data show no decrease in the particle partial pressure with $\theta_{Bn}$ for a given $M_{f}$, that is color-scaled by the right-bar. For the ACE shocks the EPAM, SWICS, and DE30 energies are summed, and for the Wind shocks the 3DP, EHSP, and SFSP are summed. Right: Ratio of downstream energy flux of energized particles (protons $+$ electrons, $(v_k\sum \rho_sh_s)^{down}$) to the upstream ram pressure flux $(\rho_g v^2 v_k/2)^{up}$. Again, no trend of decrease of energized particles contribution as $\theta_{Bn}$ increases is measured.}
    \label{fig:particle_fraction}
\end{figure}

For the ACE/Wind shocks the fractions of energy drained into the supra-thermal/energetic particle populations are summarized in Figure \ref{fig:particle_fraction}. Note that the energy fraction is defined in Figure \ref{fig:particle_fraction}, left panel, as the ratio of downstream enthalpy flux of energized particles (protons $+$ electrons) $(v_k\sum \rho_sh_s)^{down}$ to the total downstream energy flux $F^{down}_{TOT}$, differently from the efficiency used, e.g., by \cite{Ellison.Moebius:87} or \cite{Giacalone.etal:97}, that is the ratio of downstream enthalpy flux of energized particles to the upstream kinetic energy flux $(\rho_g v^2 v_k/2)^{up}$ (Figure \ref{fig:particle_fraction}, right panel). For the former definition, the fractions span from 0.48\% to 16\%, and for the latter, the fractions range from 0.75\% to 7.2\%. The reason for comparing different normalizations is that some events show a non-negligible fraction (up to $30 \%$) of incoming non-kinetic (e.g., magnetic) energy (see Figure \ref{fig:energy_fractions}); thus, the use of the kinetic energy only would lead to an underestimate of the energy available for particle acceleration. The upstream energy fluxes shown in Figure \ref{fig:RH_energy} exceed the downstream energy fluxes for all but two shocks (ACE 346/1999 and Wind 275/1998), despite being equal within $1\sigma$ (see Section \ref{sect:missing_energy}); as a result, the ratios in the right panel are typically smaller than those in the left panel. Regardless of the normalization chosen in the two panels of Figure \ref{fig:particle_fraction}, the lack of correlation of the energetic particle partial pressure with $\theta_{Bn}$ is manifest: no evidence emerges that the partial pressure of the energized particles decreases with increasing $\theta_{Bn}$. For the PSP shock, we find that energized protons (combining EPI-Lo and EPI-Hi) constitute 7.7\% of the total downstream energy flux, in agreement with our ACE and Wind results. We conclude that the energized particles  fraction is consistent with $\lesssim\!20\%$. 

Not only does the energy fraction not decrease at large $\theta_{Bn}$ for any given $M_f$; on the contrary, we found events with $\theta_{Bn} \gtrsim 70^\circ$ and energy fractions as large as $\sim\!10\%$, and having smaller relative errors than events with $\theta_{Bn} \lesssim 70^\circ$. This result confirms with high accuracy longstanding \textit{in-situ} measurements that demonstrate that ions are efficiently accelerated at quasi-perpendicular shocks \citep[e.g.,][]{Reames:12}, a result corroborated by the most recent particle-in-cell (PIC) simulations in \cite{vanMarle.etal:18}; see also the review on PIC simulation findings on the wave modes excited upstream of shocks \citep{Pohl.etal:2020}. Large-scale pre-existing upstream magnetic turbulence is suggested to enhance the ion acceleration efficiency at quasi-perpendicular shocks by numerical simulations over a broad shock parameter range \citep{Giacalone:2005,Fraschetti.Giacalone:15} and models of \textit{in-situ} STEREO measurements \citep{Fraschetti.Giacalone:20}. The results presented herein are open to such an interpretation. High-amplitude and high-frequency whistler waves excited by electrons in the shock layer \citep{Katou.Amano:19,Amano.etal:20} and traveling upstream might also drive upstream turbulence. A number of electron-scale instabilities in the high plasma beta regime of intra-cluster shocks were studied in 2D PIC \citep{Kim.etal:21}, suggesting that ion-scale waves are generated by distinct instabilities on longer time scales, partially overlapping with electron time scales.

\begin{figure}
    \gridline{\fig{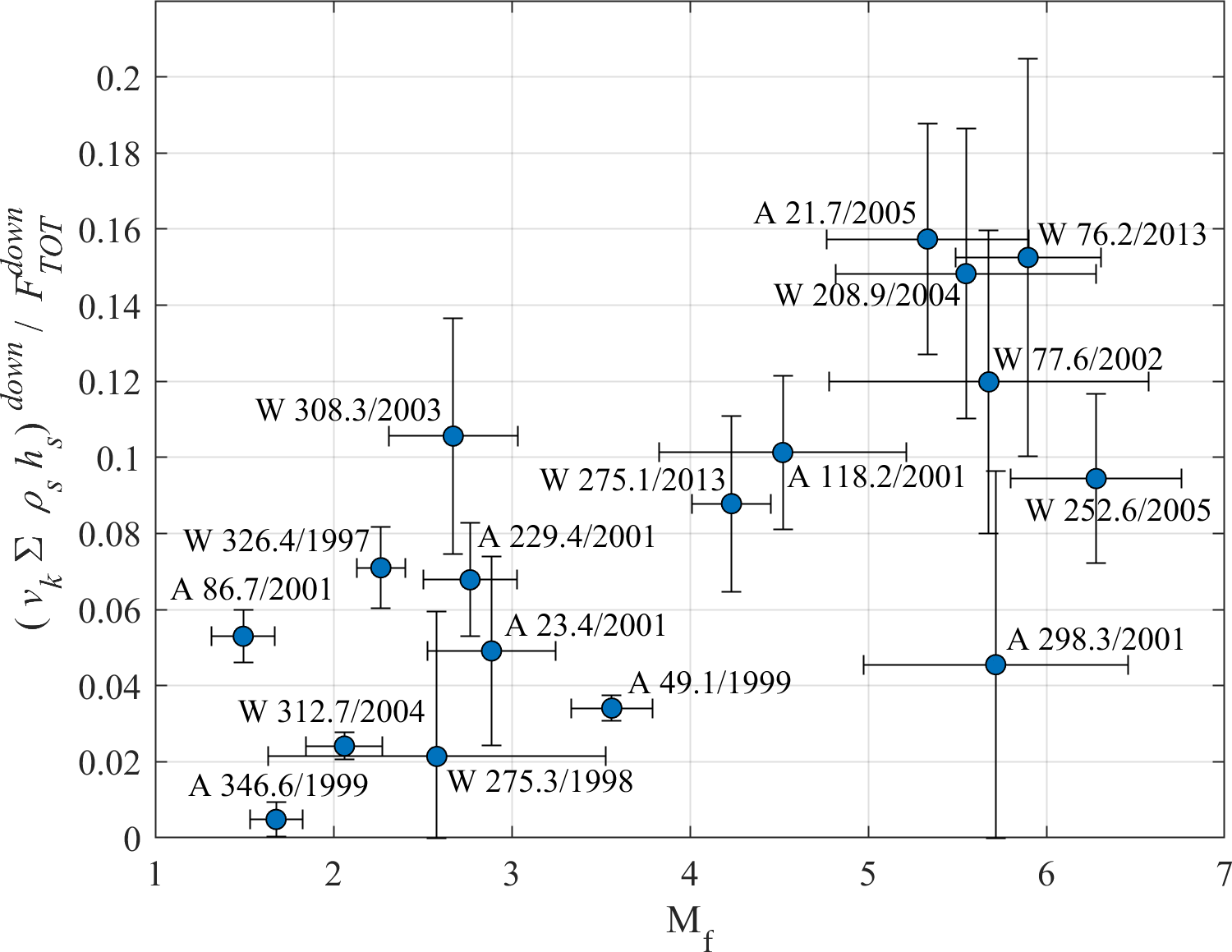}{0.5\textwidth}{}}
    \vspace{-0.5cm}
    \caption{Ratio of downstream energy flux of energized particles (protons $+$ electrons, $(v_k\sum \rho_sh_s)^{down}$) to the total downstream energy flux ($F^{down}_{TOT}$) as a function of $M_{f}$.}
    \label{fig:energy_fraction_Mach}
\end{figure}

Figure \ref{fig:energy_fraction_Mach} shows the ratio of the downstream enthalpy flux of energized particles to the total downstream enthalpy flux as a function of $M_f$. The results display a bias toward higher particle energy fractions for high-$M_{f}$ shocks. Generally, this is expected in standard DSA theory since the plasma density compression ratio, which determines the power-law slope of the high-energy tail of the distribution, increases with $M_f$ to its maximum value of 4. Moreover, the acceleration rate increases with shock speed (and, hence, $M_f$). Thus, we expect more high-energy particles with increasing  $M_{f}$. A notable exception in Figure \ref{fig:energy_fraction_Mach} is ACE 298/2001, possibly affected by one of the largest errors in the downstream energy fraction of our sample.

\begin{figure}
    \gridline{\fig{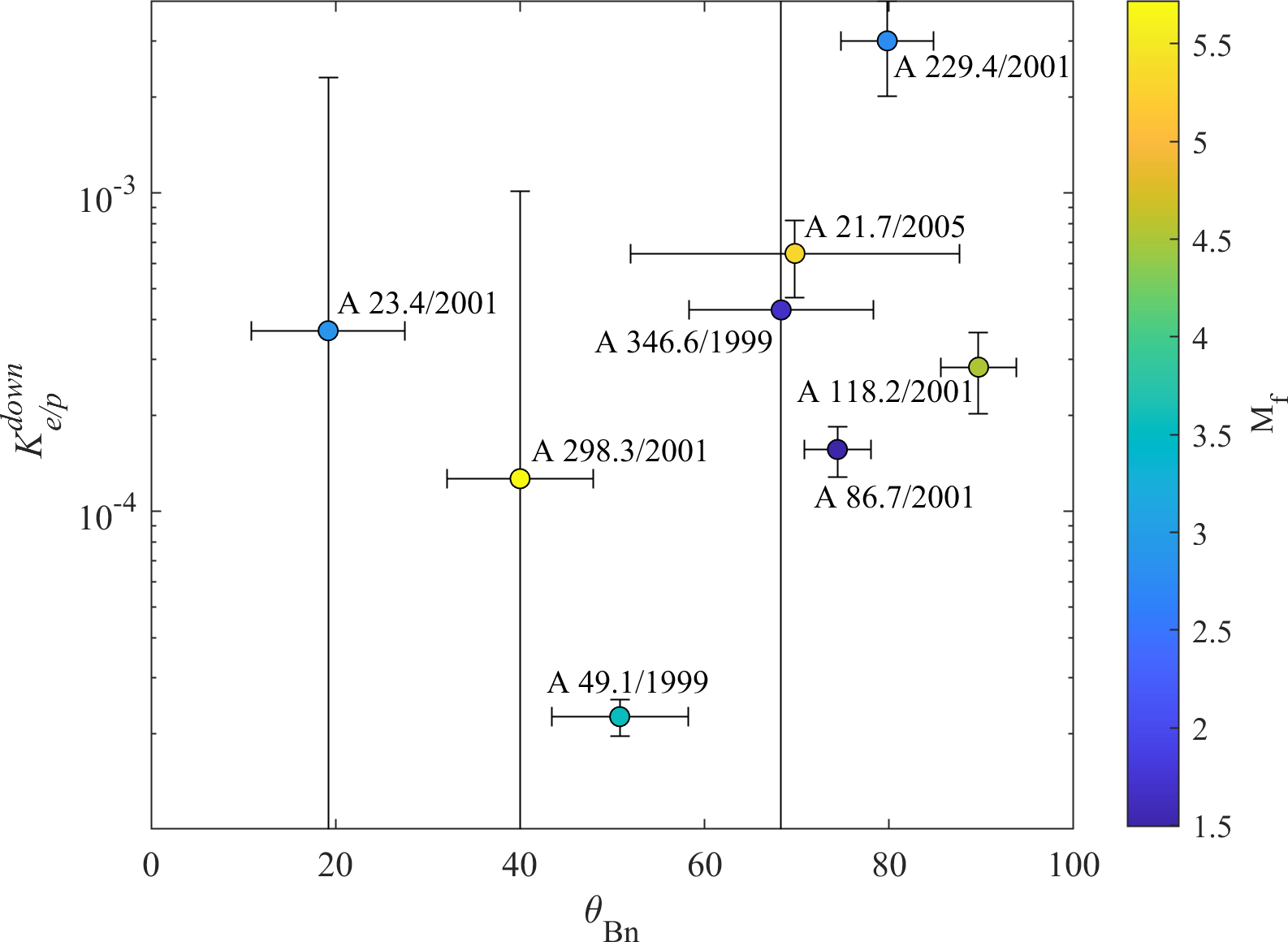}{0.49\textwidth}{}
              \fig{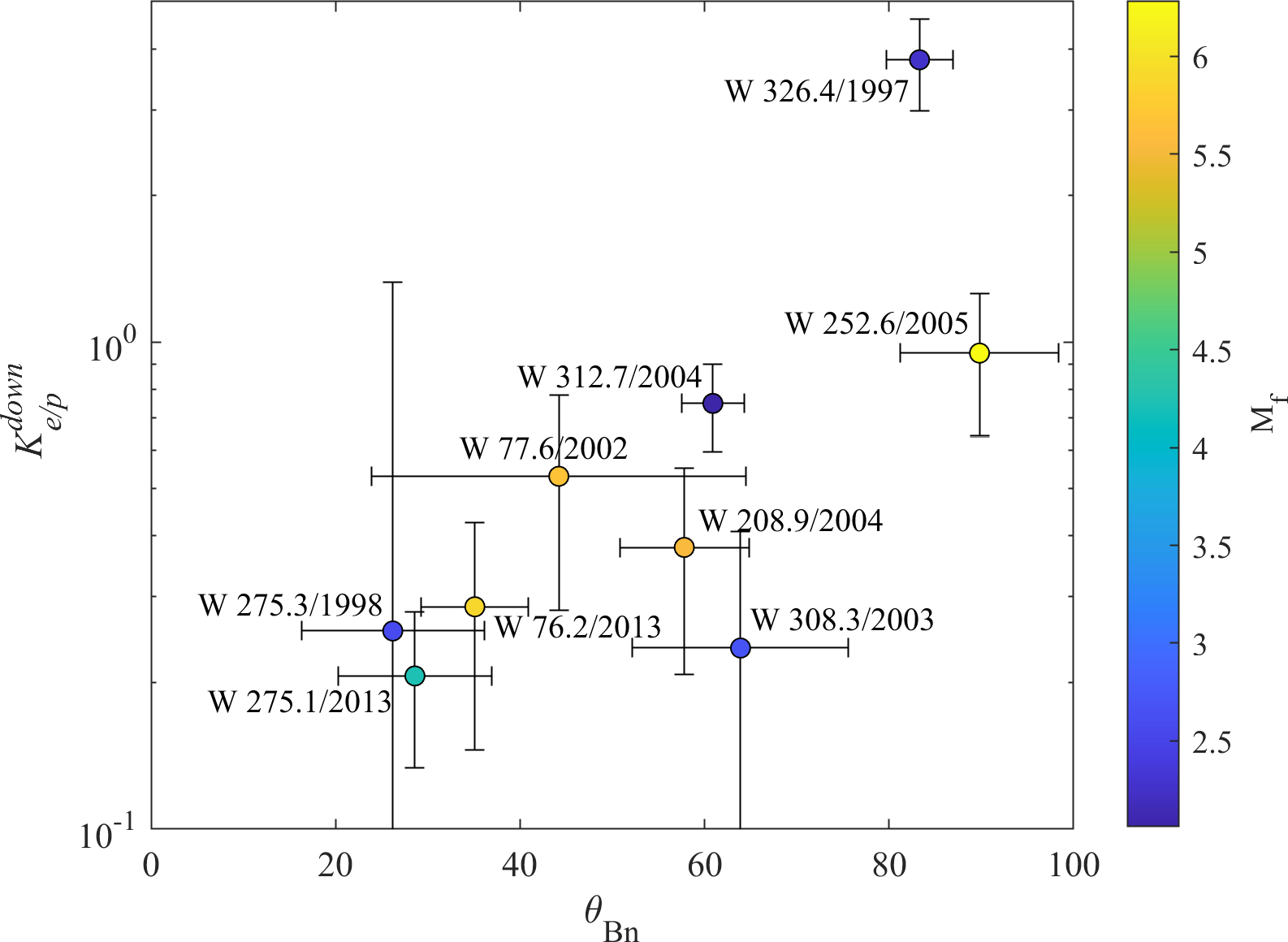}{0.49\textwidth}{}}
    \vspace{-0.5cm}
    \caption{Ratio of the downstream energized electron-to-proton energy flux $K_{e/p}^{down}$ as a function of $\theta_{Bn}$ for ACE (left panel) and Wind (right panel). The ratio for ACE shocks is roughly two to three orders of magnitude smaller than that of the Wind shocks, which might indicate that electrons with energies $<\!38\,\text{keV}$ (the lower bound of ACE/DE30) contribute a non-negligible fraction. Data below this energy are unavailable for ACE, while EHSP extends to the energy cutoff herein used for Wind.}
    \label{fig:electron_to_proton}
\end{figure}

The downstream energized electron-to-proton energy flux ratio $K_{e/p}^{down}$ is displayed in Figure \ref{fig:electron_to_proton}. Remarkably, a mismatch in the order of magnitude between ACE and Wind is found, where the former would be closer to the expected electron-to-proton post-shock temperature ratio derived from the collisionless jump conditions (given by electron-to-proton mass ratio). However, the latter is closer to the PIC-simulated high-energy tail of the spectra for quasi-parallel shocks \citep{Kato:15} and to the PAMELA (Payload for Antimatter Matter Exploration and Light-nuclei Astrophysics) $\sim\!1-600\,\text{GeV}$ electron-to-proton flux ratio cumulatively measured at Earth \citep{Adriani.etal:11a,Adriani.etal:11b}. The low value of $K_{e/p}^{down}$ for ACE events might be due to the small increase in DE30 electron intensity at the shocks not shown here, which is entirely absent for some events. Also, DE30 fluxes have an elevated instrumental background in the events considered, leading to the large error bars in Figure \ref{fig:electron_to_proton}. Contrasting all other shocks in this study, in Wind 326/1997 the energy in downstream energized electrons (EHSP) exceeds that in downstream energized (3DP) protons; for Wind 252/2005 the relative contributions match within $1\sigma$. Both of these results are likely due to missing energy components, which we discuss more in Section \ref{sect:missing_energy}.

\subsection{Time-Series Particle Energy Flux} \label{sect:time_series}
\begin{figure}
    \gridline{\fig{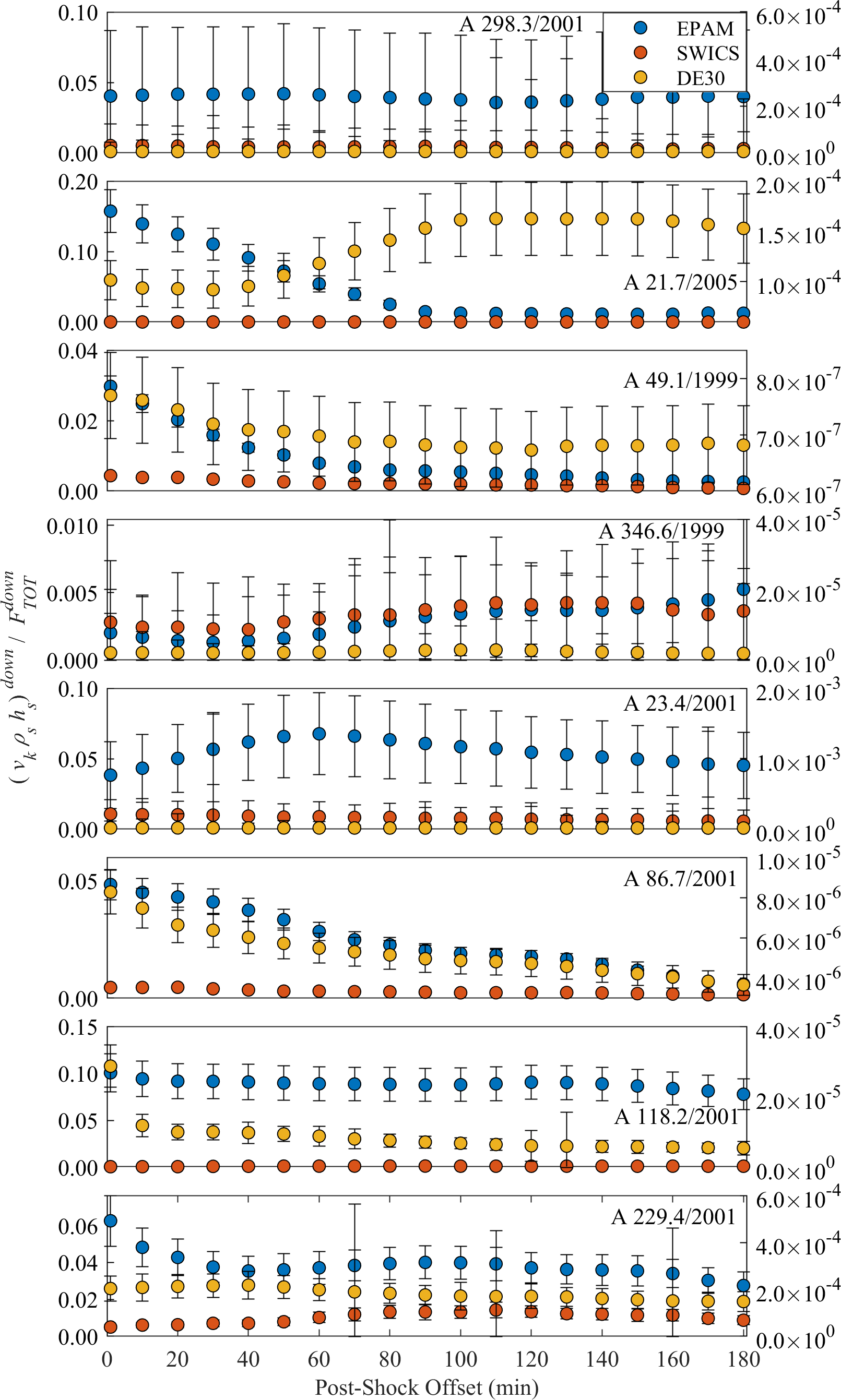}{0.49\textwidth}{}
              \fig{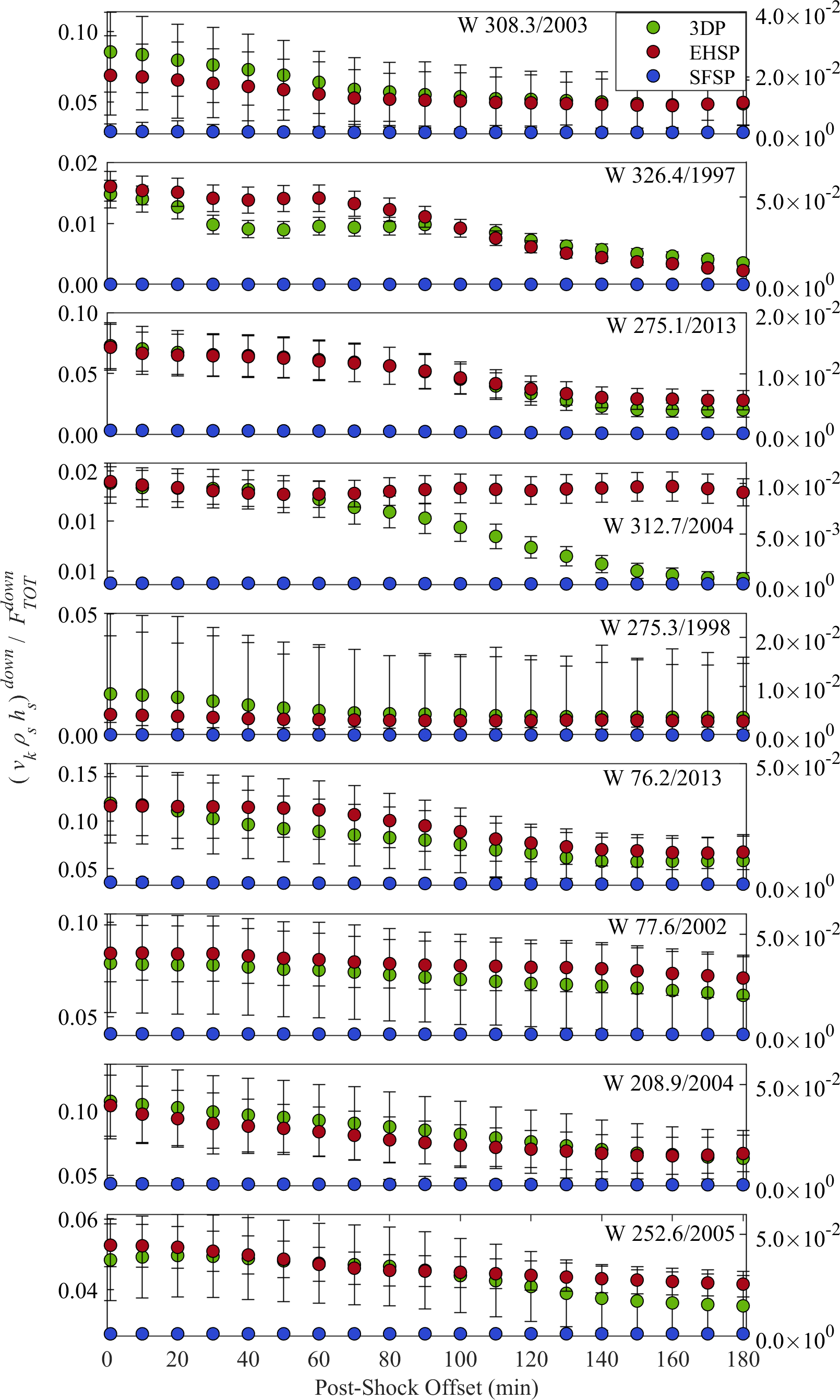}{0.49\textwidth}{}}
    \vspace{-0.5cm}
    \caption{{\bf Left:} Time dependence of the one-hour averaged downstream energy fraction of energized protons and electrons from each instrument. The averaging interval was offset from the shock time by 1 minute, and then shifted in 10-minute steps between 10 and 180 minutes (see Section \ref{sect:results}). In the ACE plots, EPAM supra-thermal ions (blue), SWICS ions (red), and DE30 electrons (yellow) are shown. The left axis scale corresponds to EPAM and SWICS, and the right to DE30. {\bf Right:} For Wind, 3DP protons (green), EHSP electrons (red), and SFSP electrons (blue) are shown. The right axis scale corresponds to 3DP, and the left to EHSP and SFSP. In most cases, there is a clear decrease in the particle energy fraction about one hour after the shock.}
    \label{fig:timeoffsetseparate}
\end{figure}

\begin{figure}
    \gridline{\fig{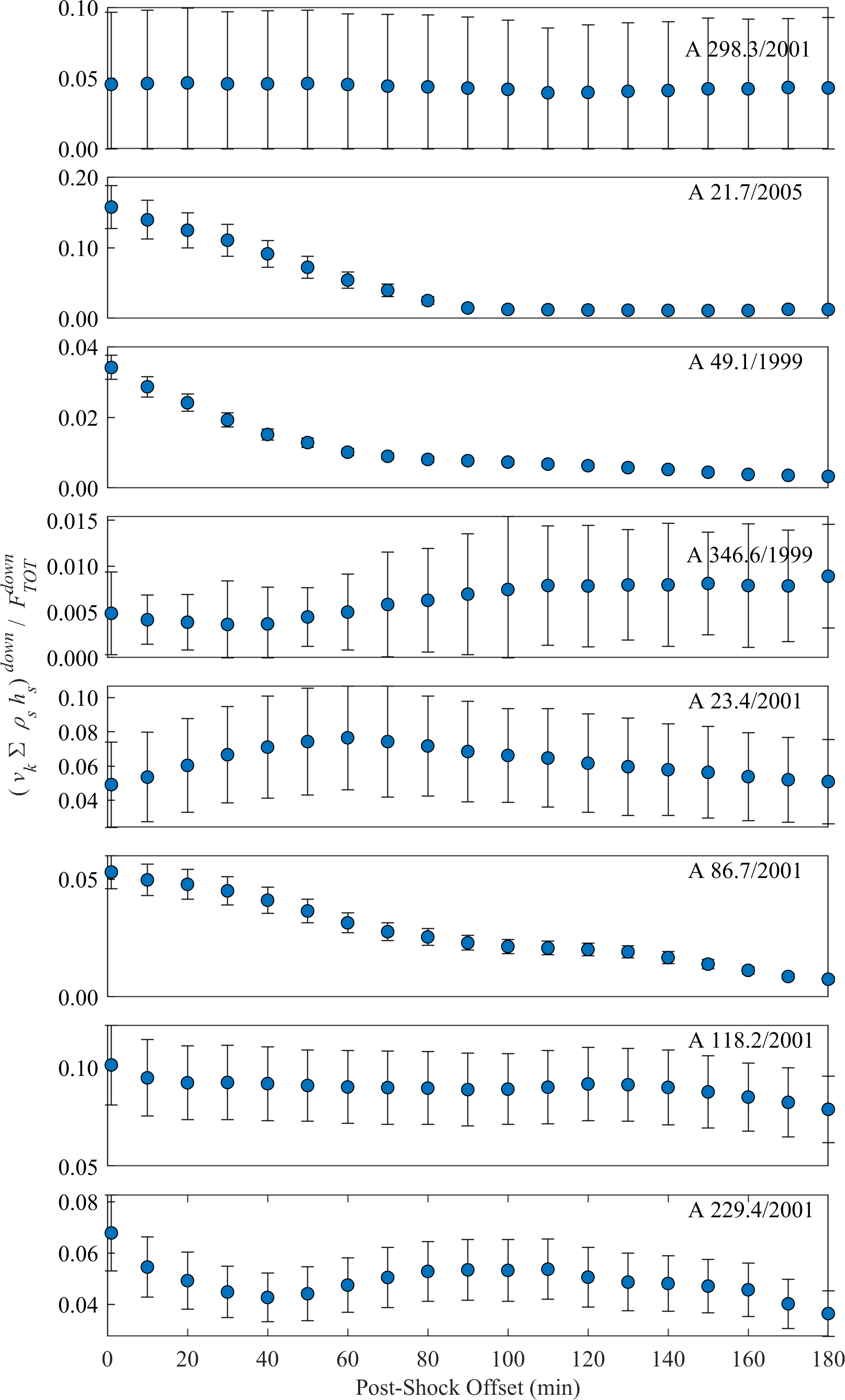}{0.49\textwidth}{}
              \fig{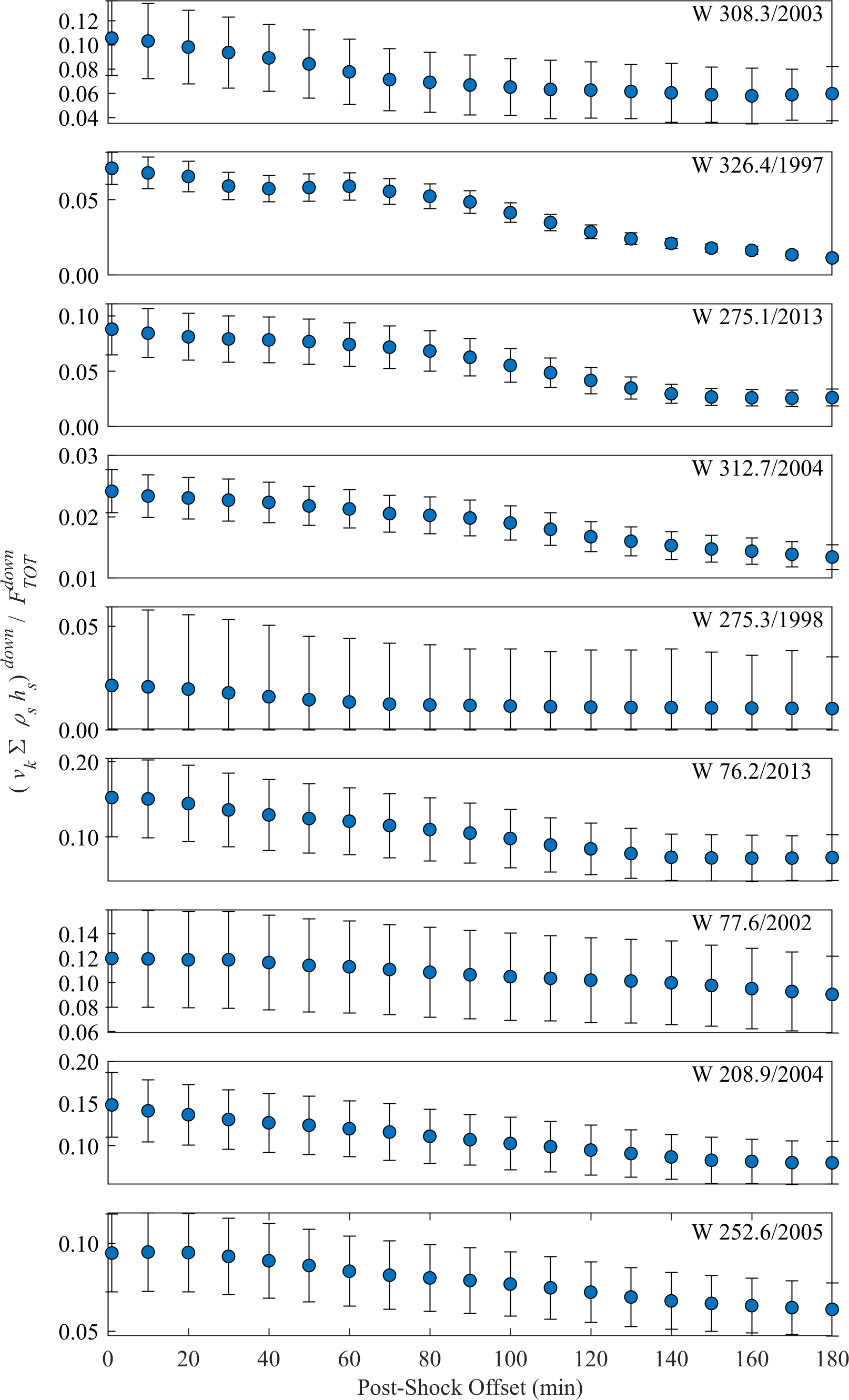}{0.49\textwidth}{}}
    \vspace{-0.5cm}
    \caption{For ACE (left) and Wind (right), change as a function of time of the one-hour averaged downstream energy fraction of energized protons and electrons summed over all instruments; the time intervals are defined as explained in Figure \ref{fig:timeoffsetseparate}. In most cases, there is a clear decrease in the particle energy fraction about 1 hour after the shock.}
    \label{fig:timeoffset}
\end{figure}

The effect of shifting the particle flux averaging time intervals after the shock time is summarized in Figures \ref{fig:timeoffsetseparate} and \ref{fig:timeoffset}. The time intervals were offset in 10-minute steps between 10 and 180 minutes after the shock time with an additional offset beginning 1 minute after the shock. A minimum of one minute was chosen to reduce the effects of strong fluctuations in particle fluxes often seen immediately after a shock, and the one-hour duration further reduces the impacts of such high-frequency variations. Since the SWICS data have a 12-minute cadence, only SWICS fluxes that fell strictly within the 1-hour windows were averaged.

In reference to Figure \ref{fig:timeoffset}, two shocks, ACE 118/2001 and 298/2001, display a constant combined proton and electron energy flux fraction up to at least three hours downstream, whereas other events exhibit a variety of time-dependencies (variations in the latter shock are likely hidden by the large errors, however). ACE 21/2005 and 49/1999 decay fastest, with the flux at 1 hour being just a third of that found immediately following the shock. ACE 86/2016, and Wind 308/2003, 326/1997, 275.1/2013, and 76/2013, drop between $15\%$ and $40\%$ over the same time. The steep drop of the downstream intensity profiles might be affected by efficient downstream particle escape \citep{Fraschetti:21}. For ACE 346/1999, however, the fraction increases with time by $\sim\!50\%$. This event precedes a much larger shock by about 0.2 days and was thus caught in the quasi-exponential upstream rise of the subsequent shock. ACE 229/2001 dips by $35\%$ at $\sim\!40$ minutes behind the shock before regaining $80\%$ of its initial flux at $\sim90$ minutes. ACE 23/2001 is maximum about 60 minutes after the shock, growing about $150\%$ relative to the value at the shock. This maximum is marked by a 300\% jump in the magnetic pressure (not shown here) and an abrupt rotation of the field, possibly the result of the spacecraft passing through a magnetic structure, e.g., a flux tube. In addition, the shock time is marked by a $\sim\!15$-minute spike \citep{Lario.etal:03} in particle intensity, twofold enhanced from the downstream intensity. This local enhancement is followed by an unusually slow rise in proton fluxes, meaning our 1-minute offset, 1-hour duration averaging interval includes the latter half of the spike but precedes the broader peak that occurs 1 hour after the shock. Such a peak might be due to downstream magnetic reconnection, invoked to explain the peaks in energetic proton fluxes about six days after Voyager 2 crossing of the heliospheric termination shock \citep{Zank.etal:2015}. No Wind shocks exhibit late downstream peaking. Since in most cases electrons constitute a significantly smaller energy fraction than protons, these trends apply to the proton-only time-series as well. DE30 electrons track EPAM protons in all discernable cases except for ACE 21/2005, where it grows to about $200\%$ its post-shock value $\sim2$ hours downstream.

The right panels of Figure \ref{fig:timeoffsetseparate} show that SFSP electrons have negligible contributions to the energy fractions at all Wind shocks. Furthermore, large errors on the associated energy fluxes make any attempts to discern trends in this energy range unfruitful. The EHSP fraction tracks closely that of the 3DP protons for all shocks except for Wind 312/2004, for which the fraction remains relatively unchanged across the three-hour interval.

The total particle partial pressures (combining electrons and protons) over all the instruments herein used are plotted in Figures \ref{fig:ace_pressure} (ACE) and \ref{fig:wind_pressure} (Wind), for both upstream and downstream plasmas. In most cases, except for ACE 49/1999 and Wind 275.1/2013, the particle partial pressure downstream exceeds its value upstream, even at the closest pre-shock sampling point. The magnetic field components and absolute value (not shown here) have no discernable upstream features that could explain the atypical behavior of these two shocks. We note two types of upstream behaviors: most events show a gradual monotonic increase over the entire 3 hours interval toward the shock time whereas two events (ACE 346/1999 and Wind 312/2004) exhibit oscillatory behavior. The peak in proton flux $\sim\!100$ minutes prior to the shock ACE 346/1999 does not correspond to evident features in the magnetic strength (e.g., a cavity) so it might be affected by other large-scale effects. The enhancement may result from magnetic connection of the spacecraft with a spatially distant region that the shock has already passed. The downstream partial pressures follow the downstream energy fractions in Figure \ref{fig:timeoffset}. The shock Wind 275.1/2013 is atypical because the partial pressure nearly flattens for the three hours before the shock and decreases at the shock. This behavior is due to an unusually large upstream flux of protons in the highest four 3DP energy bins, which exceeds that in the downstream. In other words, the passage of the shock increased the number of protons between $67.3\,\text{keV}-1.02\,\text{MeV}$ while decreasing those between $1.02\,\text{MeV}-6.76\,\text{MeV}$. No significant wave-activity is reported in the 3 hours prior to the shock time by the Wind/Magnetic Field Investigation (MFI). Similarly, we observe drop in particle pressure after the shock ACE 49/1999, but contrasting the Wind shock, the profile rises and falls quasi-exponentially upstream and downstream, and no atypical features are found in the particle fluxes.

\section{Discussion} \label{sect:discussion}

\subsection{Analysis Time Interval}
\begin{figure}
    \gridline{\fig{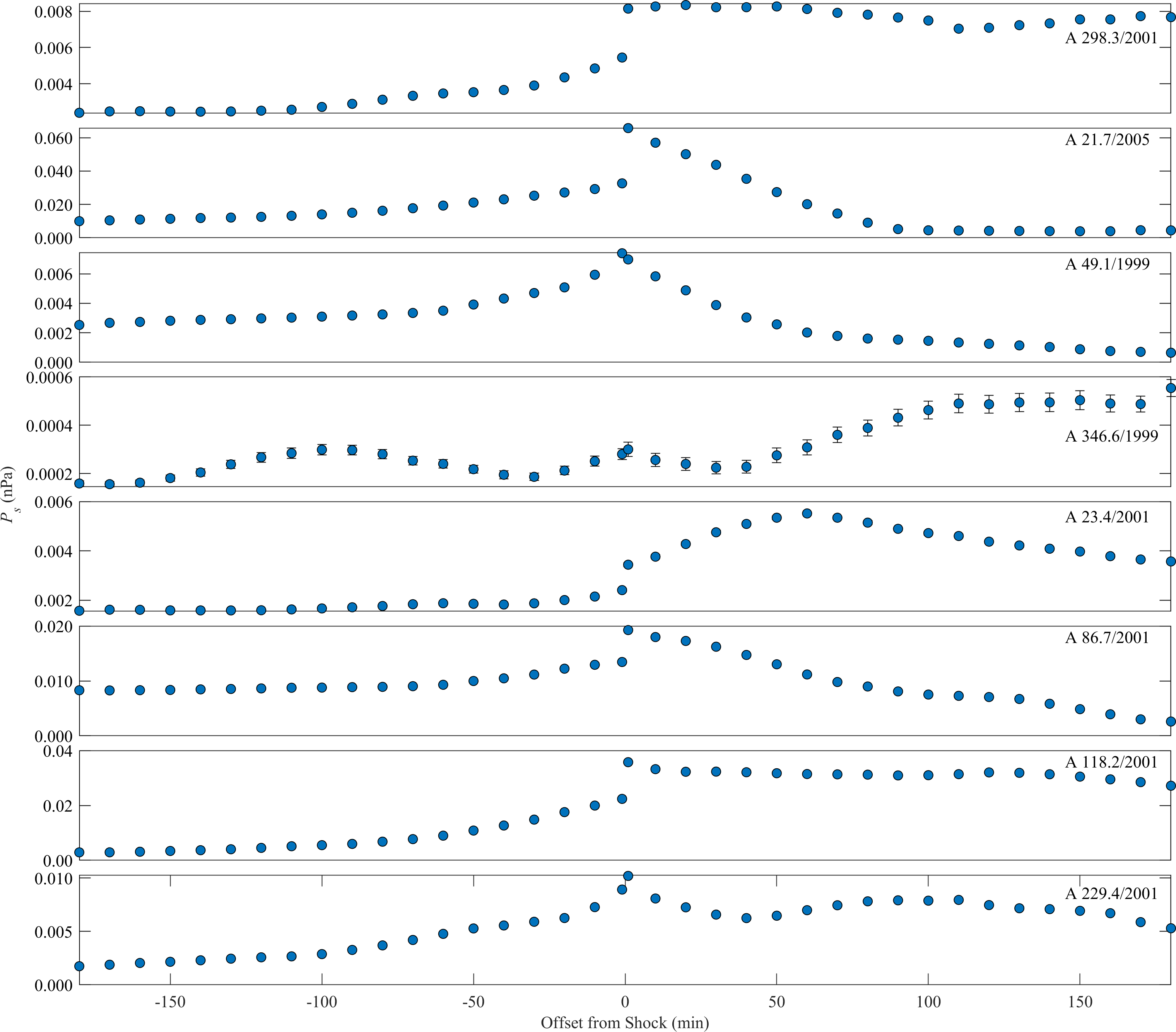}{\textwidth}{}}
    \vspace{-0.5cm}
    \caption{For each ACE shock, change as a function of time of the 1-hour average total energized particle pressure; the time intervals are as explained in Figure \ref{fig:timeoffsetseparate}, and are here combined upstream and downstream.}
     \label{fig:ace_pressure}
\end{figure}

\begin{figure}
    \gridline{\fig{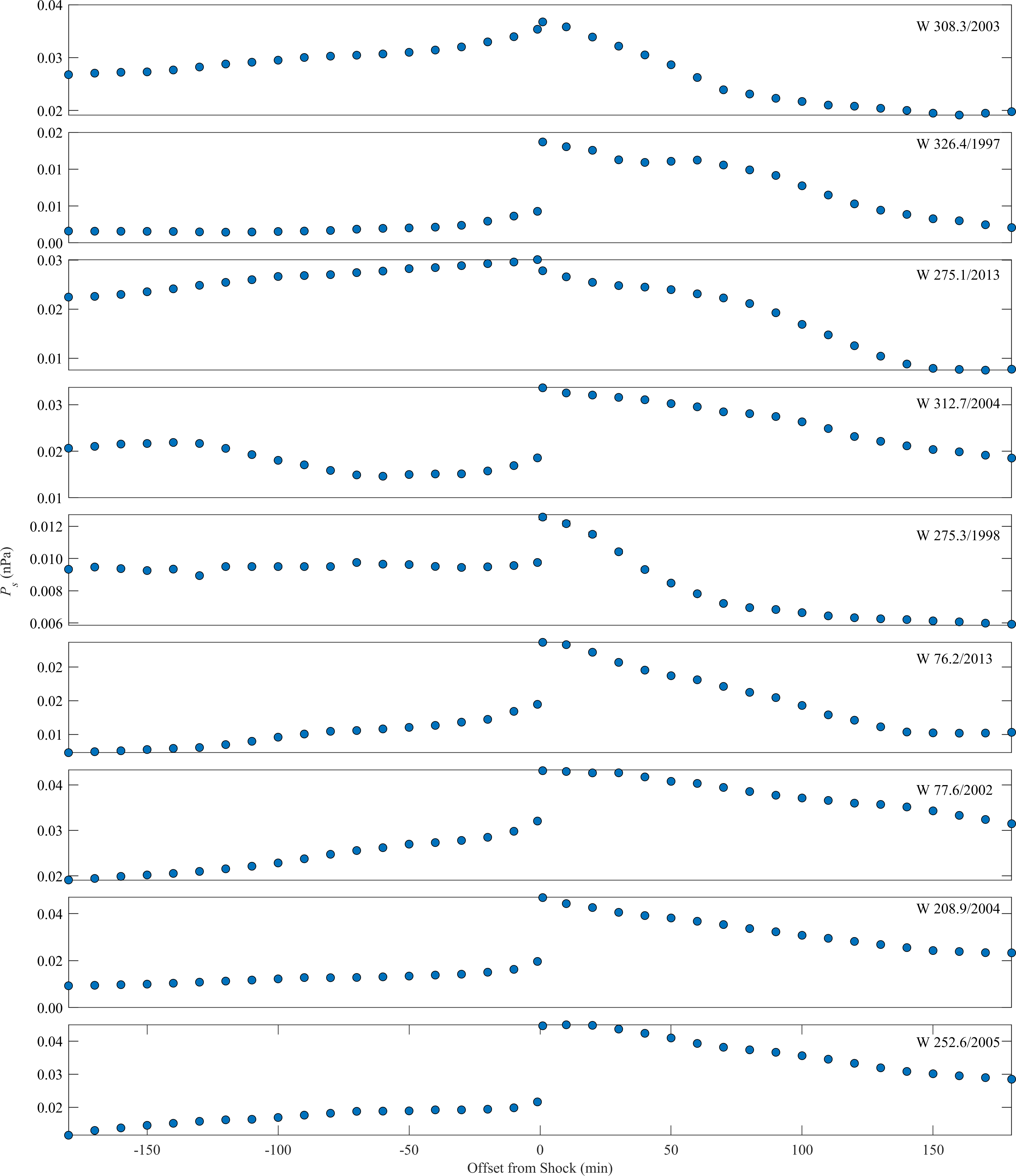}{\textwidth}{}}
    \vspace{-0.5cm}
    \caption{Same as Figure \ref{fig:ace_pressure} for Wind shocks.}
     \label{fig:wind_pressure}
\end{figure}

Earth bow shock crossings by the Magnetospheric Multiscale (MMS) spacecraft were used by \cite{Hanson.etal20} to minimize the error in the RH jump conditions: the upstream and downstream time windows were varied to determine shock parameters including $\theta_{Bn}$, the shock normal, and the shock speed with a very high time resolution. The change in the upstream and downstream shock parameters was found to be modest within the selected time-window (a $1$ minute window shifted within $5-10$ minutes of the shock time, corresponding to the sampling closest to the shock in Figures \ref{fig:timeoffsetseparate}, \ref{fig:timeoffset}, \ref{fig:ace_pressure}, \ref{fig:wind_pressure} herein). Our analysis uses a determination of shock parameters over a longer time interval (c.a. $10-20$ minutes via the CfA database) and applies a larger window (1 hour) out to much larger distances from the shock (3 hours) than MMS in order to capture the contribution of energized particles streaming away from the shock; the energetic particle intensity profiles are observed to rise over the far upstream background as early as three hours before the shock for the quasi-parallel geometry. Thus, the determination of energy conservation performed herein relies on stable parameters determined from 1 minute to 3 hours after the shock time. The downstream time window approaches the shock as close as possible to capture the freshly energized particles.

The effect of large-amplitude fluctuations in the particle intensity immediately past the shock, present in some events, on the calculated partial pressure and the energy conservation within $1\sigma$ errors is smeared out by the long time interval used in our analysis. In most cases the upstream partial pressure of energized particles immediately preceding the shock does not exceed $70\%$ of that immediately downstream.

\subsection{Missing Energy Components \label{sect:missing_energy}}
The upstream and downstream energy fluxes, including the energized particle pressure only in the downstream region, for 11 of the 17 ACE/Wind shocks in our sample, match within $1\sigma$, and all but one (ACE 21/2005) match within $2\sigma$. Moreover, for all but two shocks (ACE 346/1999 and Wind 275/1998), the upstream energy flux exceeds that downstream (although still within $1\sigma$ errors), implying missing energy components. Indeed, the Wind 3DP instrument ranges from a minimum energy of $67.3\,\text{keV}$, hence the proton energy ranges from $10.7\,\text{keV}-67.3\,\text{keV}$ (for the slowest shock) to  $57.4\,\text{keV}-67.3\,\text{keV}$ (for the fastest shock) are unaccounted for. As we mentioned above, the PESA-Hi fluxes, which range from $400\,\text{eV}-19.1\,\text{keV}$ and therefore would reduce the range of missing energies, are not included in the ion partial pressure due to large errors. Since energy spectra generally increase towards lower energies, the contribution from this energy range is expected to be non-negligible, thereby explaining at least part of the unaccounted downstream energy. 

\subsection{Electron-to-Proton Ratio \label{sect:electron_to_proton}}

The $K_{e/p}^{down}$ values for the ACE shocks ($\lesssim\!10^{-3}$) are likely to be a considerable underestimate due to the lack of electron data below $38\,\text{keV}$. Conversely, for electrons the Wind/EHSP data product extends down to $136\,\text{eV}$, but protons below $67.3\,\text{keV}$ are unaccounted for; whereas SWICS covers this energy range for ACE. Certainly a non-optimized calibration of the electron and proton instrument responses can also lead to a mismatch between the $K_{e/p}^{down}$ inferred from ACE and Wind measurements. The lack of lower-energy protons (combined with the full electron energy coverage) for Wind might explain $K_{e/p}^{down}>1$ in Wind 326/1997, and $K_{e/p}^{down}\sim1$ for Wind 252/2005. For all Wind shocks, the ratio is surprisingly close to equipartition and as expected trends upward with magnetic obliquity: even in the supra-thermal range, scatter-free electrons outrun and repeatedly cross the shock, thereby being energized by following meandering field lines as pointed out in \cite{Jokipii.Giacalone:07}.

As for $K_{e/p}^{down}$ exceeding unity in the highly perpendicular event Wind 326/1997 ($\theta_{Bn} = 83.3^\circ$), an energy transfer from protons to electrons can be ruled out due to the low collisionality of the system: the collisional frequency for energy transfer of a $100\,\text{eV}$ electron over the thermal protons is $\nu_{ep} \sim 3.9 \times 10^{-6} (n_p/\text{cm}^{-3}) \Lambda \, ({\epsilon/\text{eV}})^{-3/2}\,\text{sec}^{-1}$ that for $\Lambda \sim 20$ and $n_p \sim 3\,\text{cm}^{-3}$ (at $1\,\text{au}$) yields a time scale ($\sim 2$ months) much longer than a Sun-to-Earth shock lifetime. An alternative explanation could be efficient electron upstream pre-heating due to trapping by a two-stream instability between incoming electrons and reflected ions \citep{Scholer.etal:03}. Electron heating can also arise, only at oblique shocks, from energy transfer between specularly reflected ions and field-aligned electrons via resonant interaction with lower hybrid waves, as measured in laboratory experiments \citep{Rigby.etal:18}. The necessary wave-particle analysis is beyond the scope of this work. Despite focusing on the peak of the velocity distribution, i.e. orders of magnitude below the range explored herein, \cite{Wilson.etal:2020} found that, for 52 Wind IPs, the absolute value of, and change in, the thermal/supra-thermal electron pressure upstream/downstream is often comparable to or larger than that of ions. We show that the near-to-equipartition regime extends to supra-thermal/energetic particle regime, emphasizing the importance of electrons in the shock dynamics, and the necessity of fully kinetic numerical approach.

The values of $K_{e/p}^{down}$ for Wind shocks in Figure \ref{fig:electron_to_proton} are also one to three orders of magnitude higher than those estimated for high Mach number SNR shocks ($\sim\!10^{-4}-10^{-2}$), as deduced from galactic cosmic ray fluxes measured at Earth, multiband x-ray/gamma-ray spectra of young SNR, and simulations \citep[e.g.][]{Berezhko.etal:2009,Morlino.etal:2009,Yuan.etal:2012}. PIC simulations of SNR shocks suggest that, at high obliquity, electrons drive upstream turbulence prior to ions significant acceleration: again, fully kinetic simulations are needed to determine ion energy spectra \citep[]{Bohdan.etal:20,Kim.etal:21}. We note that low Mach number ($\sim\!2-4$) shocks in galaxy clusters \citep{Vazza.etal:2015} are also consistent with the near-equipartitioning of (relativistic) electrons and protons or electron acceleration efficiency exceeding protons ($K_{e/p}^{down}\sim\!1-100$).

\section{Conclusions \label{sect:conclusions}}

We have studied particle acceleration at $17$ interplanetary shocks crossed at $1\,\text{au}$ by the ACE and Wind spacecraft and $1$ shock crossed by PSP at $0.8\,\text{au}$. We have determined that electrons and protons energized by shocks at $1\,\text{au}$ typically drain up to $10\%$ of the upstream kinetic energy flux or up to $16\%$ of the total downstream energy density, consistent with previous estimates inferred for shocks associated with energetic CMEs \citep[e.g.][]{Mewaldt.etal:08,Emslie.etal:12,Aschwanden.etal:17} or high Mach number SNR shocks \citep[e.g.][]{Slane.etal:2014,Hovey.etal:15,Bamba.etal:18}. The presumably strong shock crossed by PSP at $0.8\,\text{au}$ is consistent with this estimate. 

We have found no significant correlation between the energy fraction in energized particles and $\theta_{Bn}$ for each given $M_f\lesssim7$, in agreement with recent combined MHD-PIC simulations \citep{vanMarle.etal:18}; the energy fraction of energized particles does not decrease with $\theta_{Bn}$ and grows with $M_f$. The events considered rule out, with unprecedented accuracy, that the acceleration efficiency, defined both with respect to the total downstream energy and to the upstream ram pressure, is significantly higher at low magnetic obliquity. Therefore, {\it in-situ} measurements show that quasi-perpendicular shocks ($\theta_{Bn} > 45^\circ$)  are not disfavored sites for efficient particle (i.e., both protons and electrons) acceleration with respect to quasi-parallel shocks ($\theta_{Bn} < 45^\circ$). 

We have estimated the energized electron-to-proton energy ratio $K_{e/p}^{down}$ in the supra-thermal and energetic particles regime. For the ACE shocks, the instrumental electron energy range does not cover energies down to the DSA cutoff, yielding an underestimate of $K_{e/p}^{down}$. For Wind shocks, the near-equipartition between the electron and proton partial pressures, possibly affected by an incomplete proton energy coverage, is considerably higher than that inferred for astrophysical shocks (typically $10^{-2}$) and suggests that the role of electrons in the dynamics of shock waves and the generation of upstream turbulence cannot be neglected. The modeling of shock evolution in the inner heliosphere must include all components in the energy balance and cannot be limited to a hybrid approach, although several hybrid simulations are well-supported by {\it in-situ} measurements \citep[e.g.,][]{Giacalone.etal:97}.

The IPs analyzed in this paper were carefully selected on the basis of the errors in each term (background plasma as well as energized protons and electrons) of the RH jump conditions, resulting in a matching of upstream and downstream total energy fluxes within $1\sigma$ for 11 of the 17 analyzed shocks. The collected sample is representative of an extensive scan of the ACE/Wind databases. For the majority of other shocks in the ACE/Wind databases, over the entire $M_f$ range, the errors on the shock speed and normal direction are too large to draw a conclusive statement on the relation between the obliquity and the fraction of total incoming energy transferred into energized particles.

The time dependencies of the downstream and upstream energized particle pressures revealed a variety of behaviors. For most events, the particle pressure rises monotonically before the shock and decays monotonically downstream, as suggested by models combining acceleration and escape \citep{Fraschetti:21}. Two shocks in our sample exhibited particle pressures peaking $1-2$ hours after the shock, potentially explained by late downstream magnetic reconnection \citep{Zank.etal:2015} or the passage of the spacecraft through a magnetic flux tube. The long averaging interval used in this study (1 hour) effectively reduces the effects of fluctuations in the particle intensities, particularly those close to the shock. Due to these rapid variations, as well as events unrelated to the shock that nevertheless affect energized particle populations, surveying the particle fluxes on timescales of several hours both upstream and downstream is important to verify their source.

Possible avenues of future research include extending as much as possible the energy coverage in the particle flux data, thereby allowing for a more appropriate quantification of the total energy flux, as well as of $K_{e/p}$ and a better verification of the jump conditions. PSP and Solar Orbiter, for example, contain instruments capable of surveying broader energy ranges \citep[e.g.][]{Kasper.etal:2016,McComas.etal:2016,Owen.etal:2020,Rodriguez-Pacheco.etal:2020}. Finally, shock crossings at multiple spatial locations equidistant from the Sun, or pan-heliospheric monitoring of the time-evolution of the particle pressure at a given shock by multiple spacecraft at distinct distances from the Sun, will certainly be topics of future studies.

\begin{longrotatetable}

\begin{deluxetable*}{llllllllll}
\tablecaption{Selected Wind and ACE shock parameters relevant to our analysis, retrieved from the CfA (cfa.harvard.edu/shocks) database. Since the temperatures were not provided in the CfA database, they were calculated from the provided thermal speed $W_{i}$ using $\frac{3}{2}k_BT=\frac{1}{2}m_pW_i^2$. See Table \ref{shocktable:small} for magnetic field values. \label{shocktable:large}}
\tablewidth{0pt}
\tabletypesize{\scriptsize}
\tablehead{
    \colhead{Shock} & \colhead{M$_{f}$} & 
    \colhead{V$_{sh}$} & \colhead{\textbf{V}$_{up}^{(sc)}\cdot\hat{\textbf{n}}$} &
    \colhead{\textbf{V}$_{down}^{(sc)}\cdot\hat{\textbf{n}}$} & \colhead{n$_{up}$} & 
    \colhead{n$_{down}$} & \colhead{T$_{up}$} & 
    \colhead{T$_{down}$} & \colhead{$\theta_{Bn}$}\\
    \colhead{(Spacecraft Year/Day/UT) [doy]} & \colhead{} & \colhead{(km s$^{-1}$)} & \colhead{(km s$^{-1}$)} & 
    \colhead{(km s$^{-1}$)} & \colhead{(cm$^{-3}$)} & \colhead{(cm$^{-3}$)} &
    \colhead{($10^4$K)} & \colhead{($10^4$K)} & \colhead{}
} 
\startdata
A 2001/298/8:2:1 [298.3347] & 5.7 $\pm$ 0.7 & 477.7 $\pm$ 8.7 & 327.4 $\pm$ 3.4 & 438.0 $\pm$ 16.1 & 4.9  $\pm$ 0.6 & 18.5 $\pm$ 1.4 & 0.6 $\pm$ 0.2 & 7.6 $\pm$ 2.3 & 40.0 $\pm$ 7.9 \\ 
A 2005/21/16:47:26 [21.6996] & 5.3 $\pm$ 0.6 & 1076.6 $\pm$ 8.7 & 598.2 $\pm$ 23.6 & 899.3 $\pm$ 16.5 & 5.9  $\pm$ 0.6 & 15.9 $\pm$ 3.1 & 10.6 $\pm$ 1.2 & 39.4 $\pm$ 8.9 & 69.8 $\pm$ 17.8 \\ 
A 1999/49/2:8:50 [49.0895] & 3.6 $\pm$ 0.2 & 718.5 $\pm$ 3.7 & 398.7 $\pm$ 7.0 & 596.7 $\pm$ 10.0 & 3.9  $\pm$ 0.6 & 10.6 $\pm$ 1.0 & 6.5 $\pm$ 0.6 & 31.8 $\pm$ 3.4 & 50.8 $\pm$ 7.4 \\ 
A 1999/346/15:14:36 [346.6352] & 1.7 $\pm$ 0.1 & 521.9 $\pm$ 3.7 & 323.1 $\pm$ 15.3 & 432.5 $\pm$ 20.3 & 0.8  $\pm$ 0.1 & 1.8 $\pm$ 0.3 & 2.2 $\pm$ 0.5 & 17.2 $\pm$ 3.7 & 68.3 $\pm$ 10.0 \\ 
A 2001/23/10:6:20 [23.4211] & 2.9 $\pm$ 0.4 & 582.4 $\pm$ 16.6 & 415.0 $\pm$ 4.9 & 546.5 $\pm$ 12.0 & 2.7  $\pm$ 0.4 & 13.0 $\pm$ 0.6 & 5.0 $\pm$ 0.9 & 30.6 $\pm$ 2.0 & 19.2 $\pm$ 8.3 \\ 
A 2001/86/17:15:15 [86.7189] & 1.5 $\pm$ 0.2 & 566.3 $\pm$ 17.4 & 394.1 $\pm$ 5.9 & 486.3 $\pm$ 12.4 & 7.4  $\pm$ 0.6 & 16.0 $\pm$ 1.0 & 11.3 $\pm$ 0.7 & 19.0 $\pm$ 0.8 & 74.4 $\pm$ 3.6 \\ 
A 2001/118/4:31:58 [118.1889] & 4.5 $\pm$ 0.7 & 906.4 $\pm$ 57.8 & 456.6 $\pm$ 8.2 & 752.3 $\pm$ 10.4 & 3.5  $\pm$ 0.4 & 11.0 $\pm$ 2.5 & 3.1 $\pm$ 0.7 & 30.8 $\pm$ 4.3 & 89.7 $\pm$ 4.1 \\ 
A 2001/229/10:16:2 [229.4278] & 2.8 $\pm$ 0.3 & 503.8 $\pm$ 8.9 & 343.9 $\pm$ 4.5 & 469.9 $\pm$ 5.5 & 3.8  $\pm$ 0.5 & 17.9 $\pm$ 1.2 & 1.5 $\pm$ 0.2 & 5.3 $\pm$ 0.5 & 79.8 $\pm$ 5.0 \\ 
W 2003/308/6:46:4 [308.2820] & 2.7 $\pm$ 0.4 & 762.3 $\pm$ 18.3 & 480.1 $\pm$ 7.4 & 722.6 $\pm$ 14.5 & 2.6  $\pm$ 0.6 & 19.1 $\pm$ 1.9 & 18.8 $\pm$ 6.6 & 104.8 $\pm$ 11.3 & 63.9 $\pm$ 11.7 \\ 
W 1997/326/9:12:52 [326.3839] & 2.3 $\pm$ 0.1 & 489.6 $\pm$ 7.9 & 344.6 $\pm$ 1.9 & 432.9 $\pm$ 6.6 & 11.7  $\pm$ 0.5 & 29.0 $\pm$ 0.2 & 6.8 $\pm$ 0.4 & 24.1 $\pm$ 3.4 & 83.3 $\pm$ 3.6 \\ 
W 2013/275/1:15:49 [275.0527] & 4.2 $\pm$ 0.2 & 654.3 $\pm$ 1.5 & 371.1 $\pm$ 4.6 & 563.4 $\pm$ 9.3 & 6.5  $\pm$ 0.3 & 20.0 $\pm$ 1.4 & 8.6 $\pm$ 1.0 & 68.5 $\pm$ 21.2 & 28.6 $\pm$ 8.3 \\ 
W 2004/312/17:59:5 [312.7494] & 2.1 $\pm$ 0.2 & 751.2 $\pm$ 27.9 & 437.8 $\pm$ 9.0 & 611.2 $\pm$ 15.0 & 18.1  $\pm$ 0.8 & 41.0 $\pm$ 5.4 & 36.4 $\pm$ 3.5 & 65.3 $\pm$ 7.2 & 60.9 $\pm$ 3.4 \\ 
W 1998/275/7:6:4 [275.2959] & 2.6 $\pm$ 0.9 & 620.3 $\pm$ 77.0 & 349.7 $\pm$ 14.7 & 519.2 $\pm$ 20.8 & 2.8  $\pm$ 1.1 & 7.7 $\pm$ 1.1 & 11.0 $\pm$ 5.0 & 44.2 $\pm$ 13.2 & 26.2 $\pm$ 9.9 \\ 
W 2013/76/5:21:28 [76.2232] & 5.9 $\pm$ 0.4 & 766.6 $\pm$ 9.8 & 425.1 $\pm$ 2.9 & 639.6 $\pm$ 23.3 & 3.3  $\pm$ 0.2 & 9.4 $\pm$ 2.1 & 4.8 $\pm$ 0.8 & 49.8 $\pm$ 26.2 & 35.1 $\pm$ 5.8 \\ 
W 2002/77/13:14:4 [77.5514] & 5.7 $\pm$ 0.9 & 580.0 $\pm$ 39.7 & 310.9 $\pm$ 8.1 & 470.8 $\pm$ 7.1 & 15.5  $\pm$ 2.4 & 37.4 $\pm$ 3.4 & 4.2 $\pm$ 0.7 & 16.3 $\pm$ 4.1 & 44.2 $\pm$ 20.3 \\ 
W 2004/208/22:25:23 [208.9343] & 5.5 $\pm$ 0.7 & 1105.7 $\pm$ 50.5 & 593.5 $\pm$ 5.8 & 966.1 $\pm$ 24.0 & 2.0  $\pm$ 0.2 & 7.2 $\pm$ 1.1 & 13.3 $\pm$ 1.8 & 123.1 $\pm$ 25.2 & 57.8 $\pm$ 7.0 \\ 
W 2005/252/13:33:1 [252.5646] & 6.3 $\pm$ 0.5 & 620.3 $\pm$ 9.0 & 315.1 $\pm$ 2.9 & 478.1 $\pm$ 11.3 & 13.7  $\pm$ 1.3 & 30.1 $\pm$ 4.7 & 3.2 $\pm$ 0.2 & 17.3 $\pm$ 4.3 & 89.8 $\pm$ 8.6 \\ 
\enddata
\label{table:1}
\end{deluxetable*}
\end{longrotatetable}

\begin{longrotatetable}

\begin{deluxetable*}{lll}
\tablecaption{Asymptotic magnetic field energy densities retrieved from the CfA (cfa.harvard.edu/shocks) database. \label{shocktable:small}}
\tablewidth{0pt}
\tabletypesize{\scriptsize}
\tablehead{
    \colhead{Shock} & \colhead{$\frac{B_{up}^2}{8\pi}$} & \colhead{$\frac{B_{up}^2}{8\pi}$} \\
    \colhead{(Spacecraft Year/Day/UT) [doy]} & \colhead{($10^{-10}$erg cm$^{-3}$)} & \colhead{($10^{-10}$erg cm$^{-3}$)}
}
\startdata
A 2001/298/8:2:1 [298.3347] & 0.2 $\pm$ 0.0  & 2.1 $\pm$ 0.7 \\ 
A 2005/21/16:47:26 [21.6996] & 1.5 $\pm$ 0.8  & 12.3 $\pm$ 7.6 \\ 
A 1999/49/2:8:50 [49.0895] & 2.1 $\pm$ 0.4  & 12.9 $\pm$ 1.7 \\ 
A 1999/346/15:14:36 [346.6352] & 0.9 $\pm$ 0.2  & 4.1 $\pm$ 0.4 \\ 
A 2001/23/10:6:20 [23.4211] & 0.6 $\pm$ 0.1  & 1.2 $\pm$ 0.6 \\ 
A 2001/86/17:15:15 [86.7189] & 5.4 $\pm$ 0.5  & 21.1 $\pm$ 2.3 \\ 
A 2001/118/4:31:58 [118.1889] & 2.5 $\pm$ 0.5  & 22.5 $\pm$ 3.6 \\ 
A 2001/229/10:16:2 [229.4278] & 0.8 $\pm$ 0.1  & 13.6 $\pm$ 3.3 \\ 
W 2003/308/6:46:4 [308.2820] & 0.5 $\pm$ 0.5  & 4.1 $\pm$ 5.7 \\ 
W 1997/326/9:12:52 [326.3839] & 1.2 $\pm$ 0.2  & 5.5 $\pm$ 2.4 \\ 
W 2013/275/1:15:49 [275.0527] & 1.2 $\pm$ 0.3  & 3.5 $\pm$ 1.5 \\ 
W 2004/312/17:59:5 [312.7494] & 14.3 $\pm$ 1.7  & 75.9 $\pm$ 4.9 \\ 
W 1998/275/7:6:4 [275.2959] & 1.7 $\pm$ 0.4  & 10.4 $\pm$ 2.7 \\ 
W 2013/76/5:21:28 [76.2232] & 0.6 $\pm$ 0.1  & 1.9 $\pm$ 2.5 \\ 
W 2002/77/13:14:4 [77.5514] & 0.9 $\pm$ 0.5  & 9.6 $\pm$ 6.7 \\ 
W 2004/208/22:25:23 [208.9343] & 0.6 $\pm$ 0.1  & 10.0 $\pm$ 4.5 \\ 
W 2005/252/13:33:1 [252.5646] & 1.2 $\pm$ 0.3  & 5.4 $\pm$ 2.0 \\ 
\enddata
\label{table:2}

\end{deluxetable*}
\end{longrotatetable}

\acknowledgments
This paper uses data from the Center for Astrophysics $|$ Harvard \& Smithsonian Interplanetary Shock Database (https://lweb.cfa.harvard.edu/shocks/). We thank Dr. M. Stevens for providing us with guidance on the database, Dr. M. Pulupa for suggestions on the PSP event and Dr. L. Wilson III for help with Wind/PESA data. This work was supported through a NASA grant awarded to the Arizona/NASA Space Grant Consortium. It was also supported in part by NASA grant 80NSSC20K1283 as part of the Heliophysics System Observatory Connect Program. The material contained in this document is based upon work supported by a National Aeronautics and Space Administration (NASA) grant or cooperative agreement. Any opinions, findings, conclusions or recommendations expressed in this material are those of the author and do not necessarily reflect the views of NASA. FF was supported, in part, by NSF under grant 1850774, by NASA under Grants 80NSSC18K1213, 80NSSC20K1283 and 80NSSC21K0119. JG was supported, in part, by NASA under Grants 80NSSC18K1213, 80NSSC20K1283 and 80NSSC21K0119. D.L. acknowledges support from NASA HGI grant NNX16AF73G and the Living With a Star (LWS) programs NNH17ZDA001N-LWS and NNH19ZDA001N-LWS, as well as the Goddard Space Flight Center Heliophysics Innovation Fund (HIF) program.

\bibliography{EnCons}{}
\bibliographystyle{aasjournal}

\end{document}